\documentclass[12pt]{article}

\usepackage{longtable}
\usepackage{pdflscape}

\usepackage{natbib}

\renewenvironment{thebibliography}[1]{%
\begin{oldthebibliography}{#1}
    \setlength{\parskip}{0.00mm}
    \setlength{\itemsep}{0.00mm}}
{\end{oldthebibliography}}

\usepackage[french,english]{babel} 
\usepackage[utf8]{inputenc} 
\usepackage[T1]{fontenc} 
\usepackage{lmodern}

\usepackage{lipsum} 
\usepackage{balance} 

\usepackage{verbatim}
\usepackage{afterpage}
\usepackage{dcolumn}

\usepackage{framed}

\usepackage{amsthm,amssymb,mathrsfs,bm}
\usepackage[tbtags]{amsmath}
\usepackage{subeqnarray}
\usepackage{eurosym}

\usepackage{relsize,exscale}
\usepackage{tabularx}
\usepackage{multirow}
\usepackage{booktabs}
\usepackage{graphicx}
\usepackage{setspace}
\usepackage{lscape}
\usepackage{vmargin}
\usepackage{color}

\usepackage{array,multirow,makecell}
\setcellgapes{1pt}
\makegapedcells
\newcolumntype{R}[1]{>{\raggedleft\arraybackslash }b{#1}}
\newcolumntype{L}[1]{>{\raggedright\arraybackslash }b{#1}}
\newcolumntype{C}[1]{>{\centering\arraybackslash }b{#1}}

\definecolor{dark-blue}{RGB}{21,85,212}
\definecolor{light-blue}{RGB}{51,153,255}
\definecolor{dark-green}{RGB}{50,150,86}
\definecolor{dandelion}{RGB}{212,174,21}
\definecolor{bloud}{RGB}{212,21,21}
\definecolor{NewPurple}{RGB}{127,0,255}

\usepackage{url}
\usepackage{enumerate}
\usepackage{fancyhdr}

\usepackage{hyperref}
\hypersetup{
    colorlinks=true,
    linkcolor=blue,
    filecolor=magenta,      
    urlcolor=cyan,
    citecolor=blue,
    pdftitle={Overleaf Example},
    pdfpagemode=FullScreen,
    }

\usepackage[hang,splitrule]{footmisc} 


\usepackage[sf,bf,tiny,center]{titlesec} 
\titlelabel{\thetitle.\enspace}
\usepackage{titletoc}

\usepackage{nicefrac}

\renewcommand{\thefootnote}{\fnsymbol{footnote}}
\newcommand{\sym}[1]{#1}

\setmarginsrb{50pt}{50pt}{50pt}{70pt}{20pt}{20pt}{20pt}{30pt}

\title{Parental environment and student achievement:\\ Does a Matthew effect exist?\,\footnotemark[1]}

\author{Gaëlle Aymeric,\footnotemark[2]\ \ Emmanuelle Lavaine\,\footnotemark[3]\ \ and Brice Magdalou\,\footnotemark[4]}

\begin{document}
%\onehalfspacing
\maketitle

\footnotetext[1]{This paper is part of the research project \textit{MaDimIn} (Contract ANR-24-CE26-3823-02), from which financial support is acknowledged. This paper was written when Brice Magdalou was on a temporary CNRS research assignment at the Aix-Marseille School of Economics. We acknowledge financial support from the French government under the “France 2030” investment plan managed by the French National Research Agency Grant ANR-17-EURE-0020, and by the Excellence
Initiative of Aix-Marseille University - A*MIDEX. We are grateful to Elena B\'arcena-M\'artin, Nicolas Gravel, Markus Jäntti, Jo Thori Lind, Vito Peragine, Rafael Salas, Daniel Sant\'in and Alain Trannoy for their useful comments.}
\footnotetext[2]{Universidad Complutense de Madrid \\ E-mail: \texttt{gaymeric@ucm.es}.}

\footnotetext[3]{CEE-M, Univ. Montpellier, CNRS, INRAE, Institut Agro, Montpellier, France. \\ E-mail: \texttt{emmanuelle.lavaine@umontpellier.fr}.}
\footnotetext[4]{CNRS, AMSE, Marseille, France. \\ E-mail: \texttt{brice.magdalou@umontpellier.fr}.}

% ------------------------------------------

\renewcommand{\thefootnote}{\arabic{footnote}}

\pagestyle{fancy}
\fancyhf{}
\chead{\textsc{Parental environment and student achievement}}
\renewcommand{\headrulewidth}{0pt}
\cfoot{\thepage} 

\begin{abstract}

This paper investigates the causal impact of the parental environment on the student's academic performance in mathematics, literature and English (as a foreign language), using a new database covering all children aged 8 to 15 of the Madrid community, from 2016 to 2019. Parental environment refers here to the parents' level of education (i.e. the skills they acquired before bringing up their children), and parental investment (the effort made by parents to bring up their children). We distinguish the \textit{persistent effect} of the parental environment from the so-called \textit{Matthew effect}, which describes a possible tendency for the impact of the parental environment to increase as the child grows up. Whatever the subject (mathematics, literature or English), our results are in line with most studies concerning the persistent effect: a favourable parental environment goes hand in hand with better results for the children. As regards the Matthew effect, the results differ between subjects: while the impact of the parental environment tends to diminish from the age of 8 to 15 in mathematics, it forms a bell curve in literature (first increasing, then decreasing) and increases steadily in English. This result, which is encouraging for mathematics and even literature, confirms the social dimension involved in learning a foreign language compared to more academic subjects.

\vspace{0.4cm}
\end{abstract}

\vspace{0.3cm}
\begin{center}
`\textit{To all those who have, more will be given, and they will have an abundance; but from those who have nothing, even what they have will be taken away}'
\end{center}
\vspace{-0.5cm}
\begin{flushright}
[Matthew (the Bible 13:12, NRSV)]
\end{flushright}

%\newpage
\onehalfspacing
%\twocolumn
% -------------------------------------------
% -------------------------------------------
\section{Introduction}
\label{intro}
% -------------------------------------------
% -------------------------------------------

It is now well-documented that the cognitive and non-cognitive abilities developed in the early childhood drive the educational, social and professional success of people throughout their entire life. It is also well-recognized that the social background and the investment of parents in their own children impact ability acquisition, which partly explains the inequalities in academic performance across children. According to the Matthew effect,\footnote{The Matthew effect, a standard concept in sociology, has been popularized by \cite{Me68}. \cite{Ri10} proposes a review of applications in several (social) sciences, including education.} in many spheres of life, `the rich get richer and the poor get poorer'. In the education field, this effect describes a possible tendency of initial advantages, in early life, to accumulate through time. Whereas the \textit{persistent effect} of the parental environment on student achievement is now admitted, the theoretical and empirical literatures are more balanced on the existence of a possible \textit{Matthew effect}. This paper aims at contributing to this debate.  

Different channels can explain the impact of parental environment on children academic performance. The first is a possible \textit{intergenerational transmission of cognitive skills}, which implies that the association between parents and children abilities can be partly driven by genetic.\footnote{\cite{HJSVVW21} identify a causal connection between cognitive skills of the parents and their children, based on a Dutch survey on math and language skills. \cite{S07} uses Korean American adoptees data to show that genetic factors explain 44\% of the variation in educational attainment and 33\% of the variation in income.} A second is through the \textit{parent's level of education}. The children can benefit from the knowledge and diplomas acquired by their parents, but also from the related positive spillovers.\footnote{The survey proposed by \cite{HLP11} concludes that the estimates of the causal effect of parent's schooling on child's schooling differ across studies, but also that selection is the main component of the intergenerational association. At the opposite, by using original Finnish data, \cite{SK19} find a strong positive causal effect (of around 0.5) from parent's to child's attained years of education.}  A third channel is \textit{parental investment}, which can be a major input in the child production of skills. In that case, a distinction has to be drawn  between cognitive and non-cognitive skills. \cite{CHS10} found that the productivity of parental investment for cognitive skills is high in the early stages of education (before 6), but tends to significantly decrease after. At the opposite, the productivity of parental investment on non-cognitive skills is found to be higher at later stages. Finally, these channels can be exacerbated by a possible \textit{assortative mating} of the parents.\footnote{\cite{BCT22} find that 75\% of the correlation in education attainment between parents and their children is driven by the joint contribution of the parents (as compared to the contribution of each parent independently). \cite{EMZ19} also find that educational assortative mating has declined among college graduates in the US since the 1960s, while it has progressively increased among the low-educated (a trend also true for the other countries studied: Denmark, Germany, the United Kingdom, and Norway).} 

The literature on the technology of skills formation, initiated by \cite{CH07} and \cite{CHS10}, tends to support the hypothesis of cumulative advantages. The authors propose and estimate a model where, at each stage of the child development, the inputs and the production technology can differ. They find that \textit{self-productivity} (the stock of skills produced at one stage augment the skills attained at later stages), for both cognitive and non-cognitive skills, becomes stronger as the child becomes older. They also observe \textit{dynamic complementary} (the productivity of an investment can be raised by skills produced at previous stages), but with a decrease in substitutability between investment in one period and the existing stock of skills. Hence, it is more and more difficult to compensate for initial endowment differences, which can imply an increasing attainment gap between advantaged and disadvantaged children.\footnote{One main policy recommendation resulting from these estimations is that successful adolescent remediation strategies for disadvantaged children should focus on fostering non-cognitive skills.}
 
On the other hand, the equality of opportunity literature suggests that, in moving towards adulthood, the child is able to free herself (at least partially) from some external factors that have determined her previous achievements. In the same vain of the age of sexual consent or the age of criminal responsibility, this theory refers to what we can call an \textit{age (of consent) for responsible choices} \cite[][]{RT16, HPRU17}. In early childhood, the child cannot be held responsible for her behaviors and achievements as they result from \textit{circumstances} not under her control.\footnote{Laziness at school, for instance, might be explained by a home environment which is neither stimulating nor rewarding.} In contrast, one can assume that an adult is able to set out personal objectives and to take free and enlightened decisions (whatever her background), such as the level of effort she decides to put at work. This prerequisite is actually necessary for the existence of freedom in itself, by considering that the life trajectory is not fully deterministic. Of course the age of consent is a normative concept here, not a precise age threshold, and it is debatable to fix it before adulthood.\footnote{\cite{RT16} emphasize that it is controversial to use years of education as an effort variable (hence after the age of consent) until the end of secondary education, and consider that only tertiary education is immune to this criticism. \cite{HPRU17} fix this age between 12 and 16, and recalculate the fraction of income inequality due to circumstances in the US and the UK, by considering that all the childhood achievements before the age of consent is a circumstance.} But with this concept in hands, one can hypothesize that initial disadvantages or the parental influence can be partially mitigated, throughout schooling, by the emancipation of the child as she grows up.\footnote{By studying the academic performance of students who are the first in their family to attend university, \cite{Edetal22} establish that some non-cognitive skills such as conscientiousness or extraversion, predict academic performance almost as strongly as standardised university admissions test scores. One can assume that such skills do not necessarily result from parental investment.} 

In this paper, we investigate a rich and never exploited database on the Madrid Community (Spain) to analyse the impact of parental environment (including parental highest schooling and parental investment) on their child's academic performance in three subjects (mathematics, literature and English as foreign language) and its change at three different education grades (Grades 3, 6 and 10, respectively about 8, 11 and 15 years old), over four academic years (from 2016 to 2019). We have combined data from various sources, provided by the Ministry of Education and Research of the Community of Madrid. First, for each grade, we have the scores of the students in each subject. The scores are normalised following a method comparable to the one used by the  OECD's Programme for International Student Assessment (PISA). Then, we have the data from two questionnaires, one sent to the students and another sent to the parents. We obtain various descriptive observations for the students and the parents (gender, country of birth, \dots) as well as behavioral observations, which can be used as proxy of the parental investment and child's effort. This repeated cross-sectional dataset covers the overall Madrid Community, including private and public schools, and gathering data for more than 320,000 students.

Our empirical results contribute to the literature in several main directions. \textit{First}, through a linear regression with fixed effects, we observe that the parental environment--both parents' highest level of education and their investment--is a strong predictor of the child's score in mathematics, literature, and English. Hence, a favorable parental environment goes hand in hand with better results for the children, what we call the persistent effect. \textit{Second}, by using interaction components to observe how this influence evolves as the child progresses through school, we find mixed but informative results on the Matthew effect: whereas the parental influence on a child's score decreases from age 11 to 15 in mathematics and follows a bell curve in literature, it is continuously and significantly increasing in English. \textit{Third}, to address potential endogeneity and establish causality, we employ an instrumental variable (IV) analysis using the historical gender gap in tertiary education in the parents’ country of origin as an instrument. These IV estimates reinforce our initial findings, providing stronger causal evidences for the persistent effect and the mixed picture regarding the Matthew effect. Finally, our results also show that while a child's own effort (proxied by time spent on homework) becomes increasingly impactful on academic results as they get older in all subjects, this effect is slightly weaker in English. This result confirms the special status of second language learning in schools \cite[][]{Ga68}: The acquisition of a new language is a highly social process, determined by the environment in which the child lives, and which can be a source of important inequality of opportunity. In contrast, remediation programs for adolescent could be effective in mathematics or literature \cite[][]{Ba07,Ba08} by fostering, for instance, non-cognitive skills, which serve as catalysts for cognitive development, such as perseverance, conscientiousness, or curiosity. Such optimism is supported by the clear indication, obtained from our data, that effort is more and more successful after some age (for all the subjects). 

The rest of the paper is organized as follows. We present in Section~\ref{sec-data} our data and some descriptive results. Section~\ref{sec-OLS} investigates the impact of parental highest schooling on student achievement, on the basis of two linear regression strategies (with or without interaction components). In this section, we also analyze the impact of parental investment and child's effort. In Section~\ref{sec-IV}, we continue our investigation into the impact of parental highest schooling, using an instrumental variable approach, to avoid any potential endogeneity issues. Finally, we discuss in Section~\ref{sec-discuss} our main results, in the light of a related literature, mainly in psychology. We also present some implications in terms of educational policies. 

% -------------------------------------------
% -------------------------------------------
\section{Data}
\label{sec-data}
% -------------------------------------------
% -------------------------------------------

This paper is based on a rich database that has never been used in academic research. From 2016 to 2019, the Ministry of Education and Research of Madrid Community (Spain)\footnote{Consejeria de Educacion y Investigacion de la Comunidad de Madrid.} has organized annual exams for all the students of the community in Grade 3 (8 years old), Grade 6 (11 years old) and Grade 10 (15 years old; not assessed in 2016). In parallel with these examinations, four questionnaires were organized for the various stakeholders: one addressed to the parents, one addressed to the student (Grade 6 and Grade 10), one addressed to the school director, and one addressed to the teachers. The main aim of these questionnaires was to assess people's own feelings about the quality of the educational system, but also to evaluate people's involvement (such as the time parents devote to their child's education, or the weekly time children spend on homework). Surprisingly enough, these data have never been used in academic research up to date. 

The first contribution of the present paper is to gather a set of disparate files and documents. We obtain a unique and harmonized cross-sectional database, covering more than 320,000 students. This database has a number of advantages, over most existing databases. First, this is not a simple survey as they cover all pupils in the Madrid community (whether in private or public schools). Then, pupils have a common identifier on examinations and questionnaires, so that it is possible to combine quantitative data on academic performance with more qualitative data, describing the educational environment in some detail. 

This study focuses on three subjects: mathematics, literature and English (foreign language). As with many of the world's leading education surveys,\footnote{A few examples: PISA, TALIS or PIAAC, for the OECD; TIMSS or PIRLS for the International Association for the Evaluation of Educational Achievement (IEA); TOEFL and Cambridge Certification.} the final exam score in each subject is calculated on the basis of the \textit{Item Response Theory} (IRT). That refers to a family of mathematical models that attempt to explain the relationship between a candidate's response to an item and that candidate's aptitude or skills. In this study, the \textit{Partial Credit Model} (PCM) is implemented \cite[see][]{MW97}. As with the Pisa results, the scoring is then transformed so that the mean is 500 and the standard deviation 100.

The database contains information for, approximately, 615,000 students: 230,000 in Grade 3, 240,000 in Grade 6 and 145,000 in Grade 10. If we take into account students whose parents responded to the questionnaire, we obtain 321,544 students: 145,096 students in Grade 3, 123,811 students in Grade 6 and 52,637 students in Grade 10. An important question concerns the impact of parents' level of education on their child's academic performance. We create a variables with three `homogenous' categories, based on the highest level of education observed among parents:\footnote{We could also have chosen the father's highest level, or that of the mother. In this paper, we do not investigate whether the effects of parental schooling can be explained by assortative mating, or if the partial effects of parents can be differentiated. We consider \textit{parents' education as a potential} from which children can benefit. See \cite{HLP11}, Page 3, for a detailed discussion on this issue.} before Grade 11, Grade 12/Vocational Training/Short-Cycle Tertiary Education and Bachelor/Master/Doctorate. Respectively, they are coded from 0 to 2 and correspond, according to the \textit{International Standard Classification of Education} (ISCED 2011), respectively to Levels 0 to 2, Levels 3 to 5 and Levels 6 to 8. The rest of the data is described in Tables~\ref{var-desc} to~\ref{stud-dist}, in Appendix.

We illustrate in Figure~\ref{Math-Scores} the impact of the parents' highest level of education on child's mathematics scores, for Grades 3 and 10. We plot the cumulative distribution functions of the scores, conditional on the parents (highest level of schooling) group. 
\begin{figure}[!htp]
\centering
\caption{\label{Math-Scores} CDFs of mathematics scores according to parents' highest level of education} \vspace{0.2cm}
\small
\begin{tabular}{cc}
\includegraphics[width=8cm]{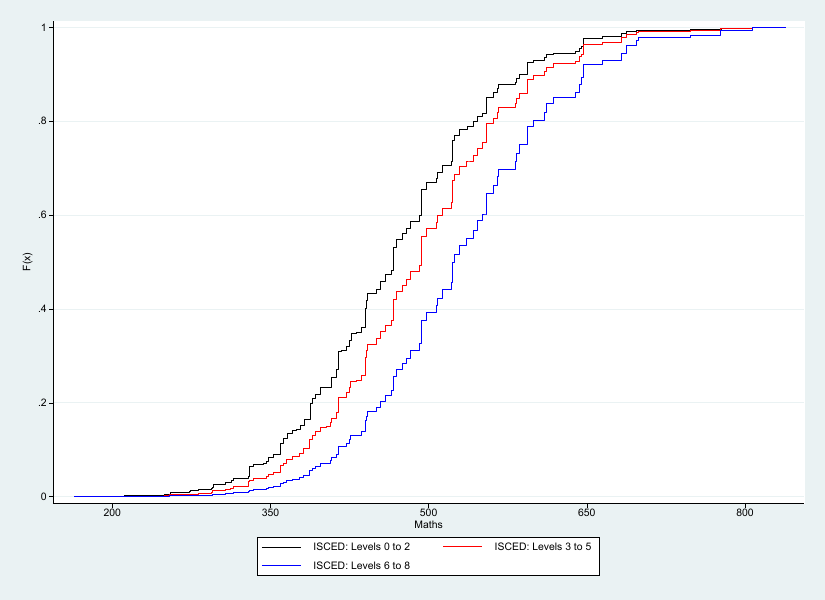} & %
\includegraphics[width=8cm]{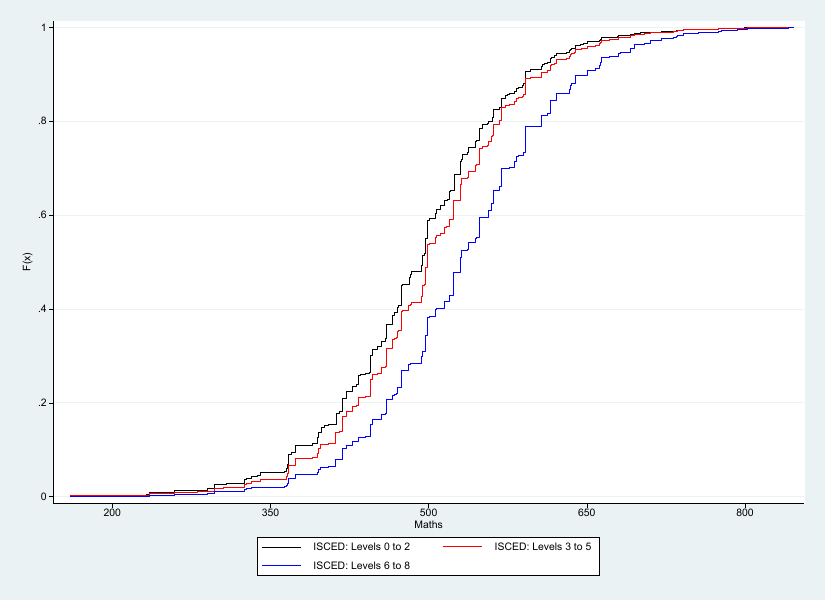} \\ 
Grade 3 & Grade 10 \\  
\end{tabular}
\end{figure} 
For each grade, we can see that the cumulative distribution functions are ordered in the sense of \textit{first order stochastic dominance}: For any given score, the probability of having a score higher than it, is all the greater the higher the parent group. This pattern holds true for mathematics, but it also applies to all subjects and all grades in our data (discussed hereafter). First-order dominance is generally considered to be a clear indication of \textit{inequality of opportunity}, since academic performance depends on a dimension beyond the child's control \cite[][]{LPT09,JGM13}. This result is in line with a robust trend already observed in the literature.

The second question looks at the evolution of parental influence on child's results, throughout schooling. In Figure~\ref{Math-Scores}, we see a convergence of the three conditional CDFs between Grades 3 and 10 (in mathematics), suggesting that the dependence of results on parent group decreases. However, such an observation must be treated with caution. First, the \textit{tests are of a different nature} for each grade. Then, the fact that the results are standardized (average of 500) can be misleading for comparison purposes. Finally, external factors can significantly impact the results, that can only be analyzed by econometric estimates. 

One possible bias in the comparison of the impact of parents' level of education on child's score, at different grades, may be linked to a \textit{composition effect}. For instance, the convergence of the CDFs between Grades 3 and 10 may be a consequence of the fact that, in the group of children whose parents have the lowest level of education, only the most gifted children remain represented in Grade 10. As established in Table~\ref{stud-dist} in Appendix, we do not observe, for each year and each grade, any significant difference in the proportions that each group represents: The proportions are all close to those obtained at global level, i.e. 9.5\%, 31.5\%, and 59\% for parents with ISCED levels, respectively,  0--2, 3--5, and 6--8. We can therefore consider that our data do not suffer from this compositional bias. 

The equivalent of Figure~\ref{Math-Scores} in literature and English is provided, respectively, in Figures~\ref{Lit-Scores} and~\ref{Eng-Scores} in Appendix. Whereas the convergence of CDFs is evident in mathematics and literature between Grades 3 and 10, this is not the case for English: The inequality of opportunity observed at Grade 3 seems to increase at Grade 10. This finding is confirmed by the econometric analysis presented in the following sections.

% -------------------------------------------
% -------------------------------------------
\section{Initial estimation strategy and results}
\label{sec-OLS}
% -------------------------------------------
% -------------------------------------------

% -------------------------------------------
\subsection{Impact of parent's highest level of education}
\label{sec-educ}
% -------------------------------------------

In this section we regress students' score in the three subjects under consideration (three regressions), focusing on the impact of parents' highest level of education. The score of student $i$ is denoted $y_{its}$, where $t$ is the year (from 2016 to 2019) and $s$ the school attended.  As described earlier, parents' highest level of education is modeled by an ordered categorical variable with 3 possible values, hereafter referred as \textit{parent groups}. We introduce independent dummy variables ($\textnormal{\textit{p}}_j$ with $j=0,1,2$) that allow us to compare the impact of parent groups two by two, with the lowest group ($j=0$) as the reference value.

\begin{figure}[!htp]
\centering
\caption{\label{Global} Global impact of parents’ highest level of education on student's score} \vspace{0.2cm}
\includegraphics[width=18cm]{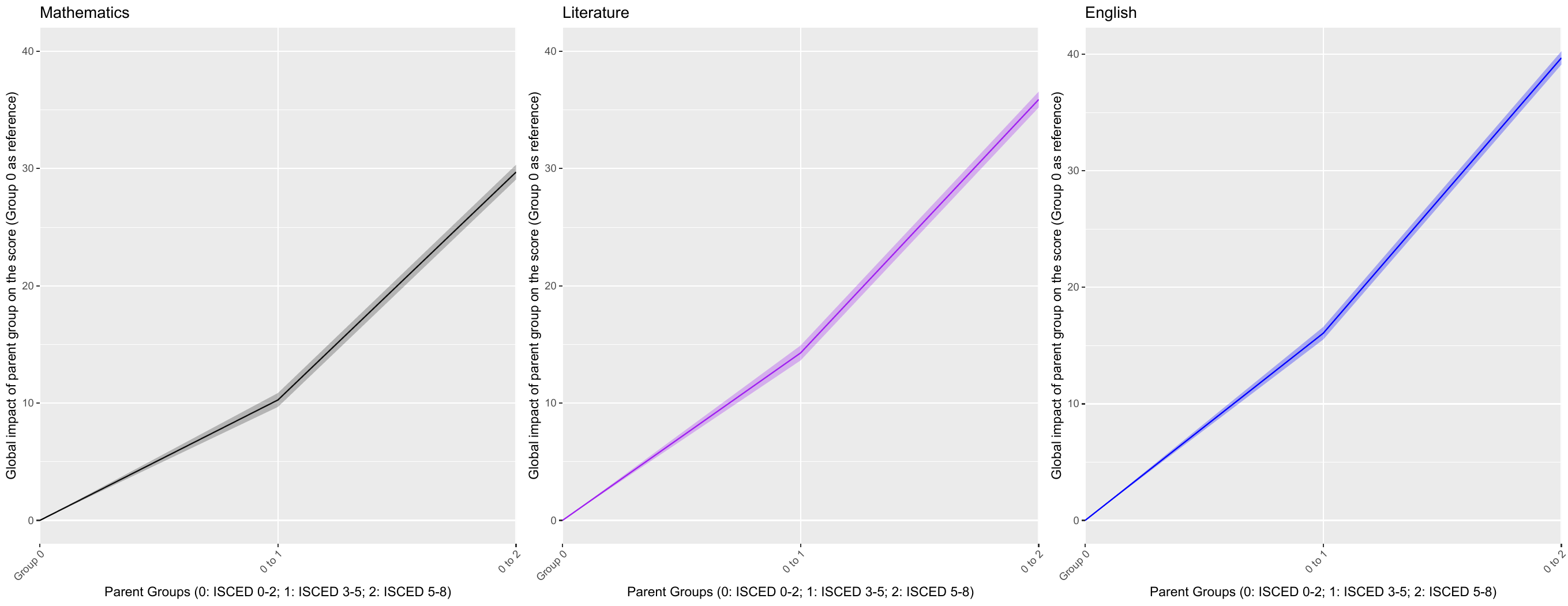}\\ 
\raggedright
\footnotesize{\textit{Reading}: In mathematics, moving from Parent Group $0$ (Education ISCED 0--2) to Parent Group $1$ (Education ISCED 3--5) increases the student's score by about 10 points (which represents 2\% of the normalised average score and 10\% of the normalised standard deviation, respectively equal to 500 and 100). $\pm$ standard deviation around the curve.}
\end{figure}

We also introduce two groups of control variables, one for the \textit{student characteristics} (each denoted $s_k$) and another for the \textit{households characteristics} (each denoted $h_k$).\footnote{Student characteristics include country of birth, number of days a week spent on homework, and gender. Household characteristics include country of birth of both parent, the frequency of use of books/computers/internet at home, the number of books at home, the employment status of both parents, the frequency with which parents talk about school subjects/teach homework/help with homework/check homework with the child.}
For each subject, we make an overall estimate of the student's score for all academic years and all grades. Although the tests are common to all schools in the Community of Madrid over the whole period, results may vary in time (from one year to another), and in space (from one school to another, particularly between public and private schools). With the aim of controlling these dimensions, we introduce \textit{two fixed effects}, one for the academic year ($a_t$) and one for the identifier of the school $s$ where the student $i$ is registered ($b_s$).\footnote{At this stage, we do not introduce a fixed effect for the grade because the mean score, whatever the grade, is always normalised at 500 at the population level (and the standard deviation at 100). Even if there is a small variation in average scores between subjects and grades in our samples, they remain very close to 500. Notice that the constants estimated in regressions~(\ref{first-reg}) is close to this value (see results tables). } 
One obtains the following regression:
\begin{equation} \label{first-reg}
y_{its} = \alpha_0 + \textstyle\sum_{j=1,2} \alpha_j\,p_{ji} + \textstyle\sum_{k} \beta_k\,s_{ki} + \textstyle\sum_{k} \gamma_k\,h_{ki} + a_t + b_s + \epsilon_{its}\,.
\end{equation}
This first regression, for each subject, isolates the influence of parents' highest level of education on student's score, without distinguishing the grade. The results are shown in Table~\ref{maths-first} for mathematics, Table~\ref{lit-first} for literature and Table~\ref{eng-first} for English. Whatever the regression specification (including or not the control variables), and for each subject, the results are the same and robust: moving from one parent group to a higher one, significantly increases student's score. For instance, in mathematics, moving from Parent Group $0$ (ISCED 0 to 2) to Parent Group $2$ (ISCED 6 to 8) increases the score by 29.7 points on average when all control variables are included (regression 3 in Table~\ref{maths-first}). The effect is larger in literature than in mathematics, and in English than in literature. We summarise these results in Figure~\ref{Global} (estimates with all control variables, also plotting standard deviation). This result confirms a trend already widely observed in the literature, namely that the parents' level of education has a major influence on the child's academic performance, whatever the subject. 

A more complex question (at the heart of this paper) concerns the evolution of this impact over time, throughout children's education. To this end, we supplement the previous regression with interaction components between parental highest education and grade level of the student (Grade 3, Grade 6 or Grade 10). Precisely, we compare the grades two by two (3 vs. 6, 6 vs. 10 and 3 vs. 10), running a regression for each possible grades pair $(u,v)$, where $u<v$. In each of these regressions, we retain only the observations of students' scores for grades $u$ and $v$ (excluding the third grade). With $g$ indicating the grade, we introduce a dummy variable $I(u,v)$ which takes the value $0$ if $g=u$ (reference grade), and $1$ if $g=v$. When we compare the grades $u$ and $v$, we obtain the following expression, with $g \in \{u,v\}$:
\begin{equation}\label{second-reg}
y_{igts}  = \alpha_0 + \textstyle\sum_{j=1,2} \alpha_j\,p_{ji} + \eta I(u,v) + \textstyle\sum_{j=1,2} \theta_j\,p_{ji} I(u,v) + \textstyle\sum_{k} \beta_k\,s_{ki} + \textstyle\sum_{k} \gamma_k\,h_{ki} + a_t + b_s + \epsilon_{igts}\,.
\end{equation}

The $\eta$ coefficient indicates the extra points obtained on average by the students in Grade $v$, compared with Grade $u$, whatever the parent group. As the average score is standardised at 500 for all grades, this coefficient is not very informative (values other than $0$ are due to sample selection). We focus on coefficients $\alpha_j$ and $\theta_j$, for $j=1,2$. By definition of $p_{ji}$ and $I(u,v)$, the variable $p_{ji} \times I(u,v)$ takes the value $1$ if and only if student~$i$ is in Grade $v$, with parents from group $j$. Since the parent reference group is the lowest ($j=0$), and the same applies to the grade (the reference is $u$, with $u<v$), the $\alpha_j$ coefficient is interpreted as the extra points obtained by the student in Grade $u$ when the parents are in group $j$ (compared with 0), and $\theta_j$ is interpreted as the marginal impact of parent's group $j$, when the student is in Grade $v$ instead of Grade~$u$. 

For each subject, we obtain 9 regressions (3 regressions, depending on whether the student and household characteristics are included or not, for each of the 3 pairs of grades compared). The results are shown in Tables~\ref{G3-G6}, \ref{G6-G10} and~\ref{G3-G10} for, respectively, the comparison of Grade 3 versus Grade 6, Grade 6 versus Grade 10, and Grade 3 versus Grade 10 (each table showing the results for the 3 subjects). As established in the regressions without interaction components, parents' highest level of education significantly increases student's score, in each subject (coefficients $\alpha_j$). If we focus on the regressions that take into account all the control variables (columns 3, 6 and 9 for each subject in, respectively, Tables~\ref{G3-G6}, \ref{G6-G10} and~\ref{G3-G10}), we can see that moving from Parent Group 0 (ISCED 0 to 2) to Group 1 (ISCED 3 to 5) significantly increases the score, between 11 and 16 points in the three subjects. From Parent Group 0 to 2 (ISCED 6 to 8), the score increases between 28 and 37 points. These last values are relatively high, in the range of a third of the normalised standard deviation at the population level (equal to 100).  

\begin{figure}[!htp]
\centering
\caption{\label{Interactions}Marginal impact of parents’ highest level of education on student's score} \vspace{0.2cm}
\includegraphics[width=18cm]{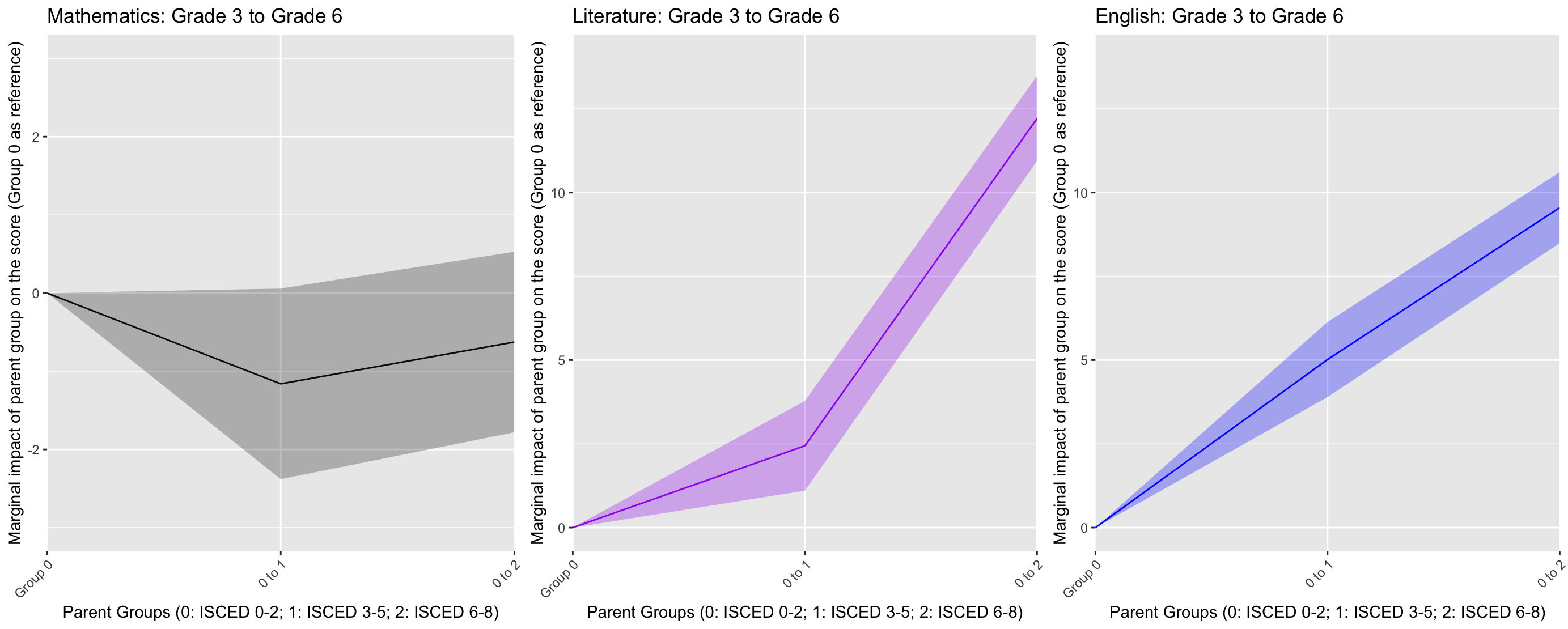}\\ 
\includegraphics[width=18cm]{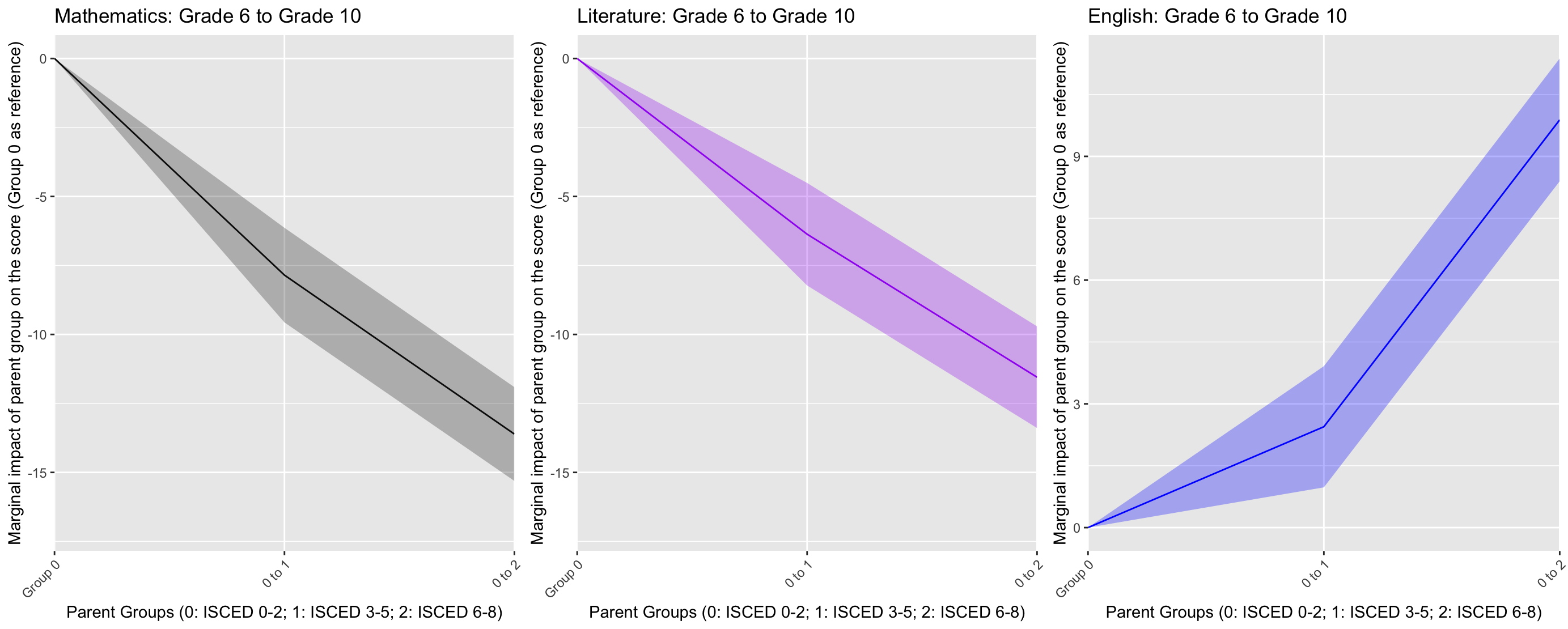}\\ 
\raggedright
\footnotesize{\textit{Reading}: Between Grade 6 and Grade 10 in mathematics, the marginal impact on the student's score, of Parent Group $1$ (ISCED 3 to 5) compared with Parent Group $0$ (Education ISCED 0--2), is reduced by about 7.5 points (which represents 1.5\% of the normalised average score and 7.5\% of the normalised standard deviation, respectively equal to 500 and 100). $\pm$ standard deviation around the curve.} 
\end{figure}

We now examine the estimates of the interaction components ($\theta_j$), again focussing on the regressions that take into account all the control variables (columns 3, 6 and 9 for each subject, of Tables~\ref{G3-G6}, \ref{G6-G10} and~\ref{G3-G10}). In mathematics, there is no significant marginal impact of parent group between Grades 3 and 6, but the impact is significant and negative between Grades 6 and 10 (and, consequently, also between Grades 3 and 10). As a result, the impact of parents' highest level of education diminishes as student progresses through grades, from Grade 6 upwards. Between Grades 6 and 10, moving from Parent Group 0 to 1 results in a 7.9 point decrease in the score gap, and moving from Parent Group~0 to 2 results in a 13.6 point decrease in the score gap. A comparable pattern applies to literature, with a small difference: Between Grades 3 and 6, the marginal impact of moving from Parent Group 0 to 2 is significantly positive (about 12.2 points). The same change in parent group, but between Grades 6 and 10, implies a gap reduction of 11.6 points. The effect is therefore  a bell-shaped curve in literature. The situation is completely different in English. Between Grades 3 and 6, and also between Grades 6 and 10, the impact of parent group increases significantly. For instance, between Grades 3 and 6, the marginal impact of moving from Parent Group 0 to 2 is around 9.5 points and, from Grade 6 to Grade 10, is about 9.9 additional points. Hence, influence of parent group is continuously and significantly increasing in English. We summarise the main results of the estimates with interaction components in Figure~\ref{Interactions} (estimates with all control variables).
%

% -------------------------------------------
\subsection{Impact of parental investment and student's effort}
\label{sec-invest}
% -------------------------------------------

In this paper, we analyse the impact of parental environment on academic performance for their child through two channels: parent's level of education (thus, skills acquired before bringing up children), and parental investment (an effort made by parents during the upbringing of their child). In the previous section, we showed that the effect of parents' education tends to decrease between Grades 6 and 10 in mathematics and literature, but increases in English. This section focuses on the second aspect, i.e. the impact of parental investment, but also on the impact of efforts made by the child in her studies. We run regressions comparable to Equation~(\ref{first-reg}), but one per subject and per grade. The results are presented in Table~\ref{parent-invest}. We focus on two explanatory dimensions for, respectively, parental investment and child's effort: `frequency parents talk to their child about school' and `days per week devoted to homework' (both are introduced in the form of dummy variables).

\begin{figure}[!htp]
\centering
\caption{\label{stud.eff}Impact of student's effort on score} \vspace{0.2cm}
\includegraphics[width=18cm]{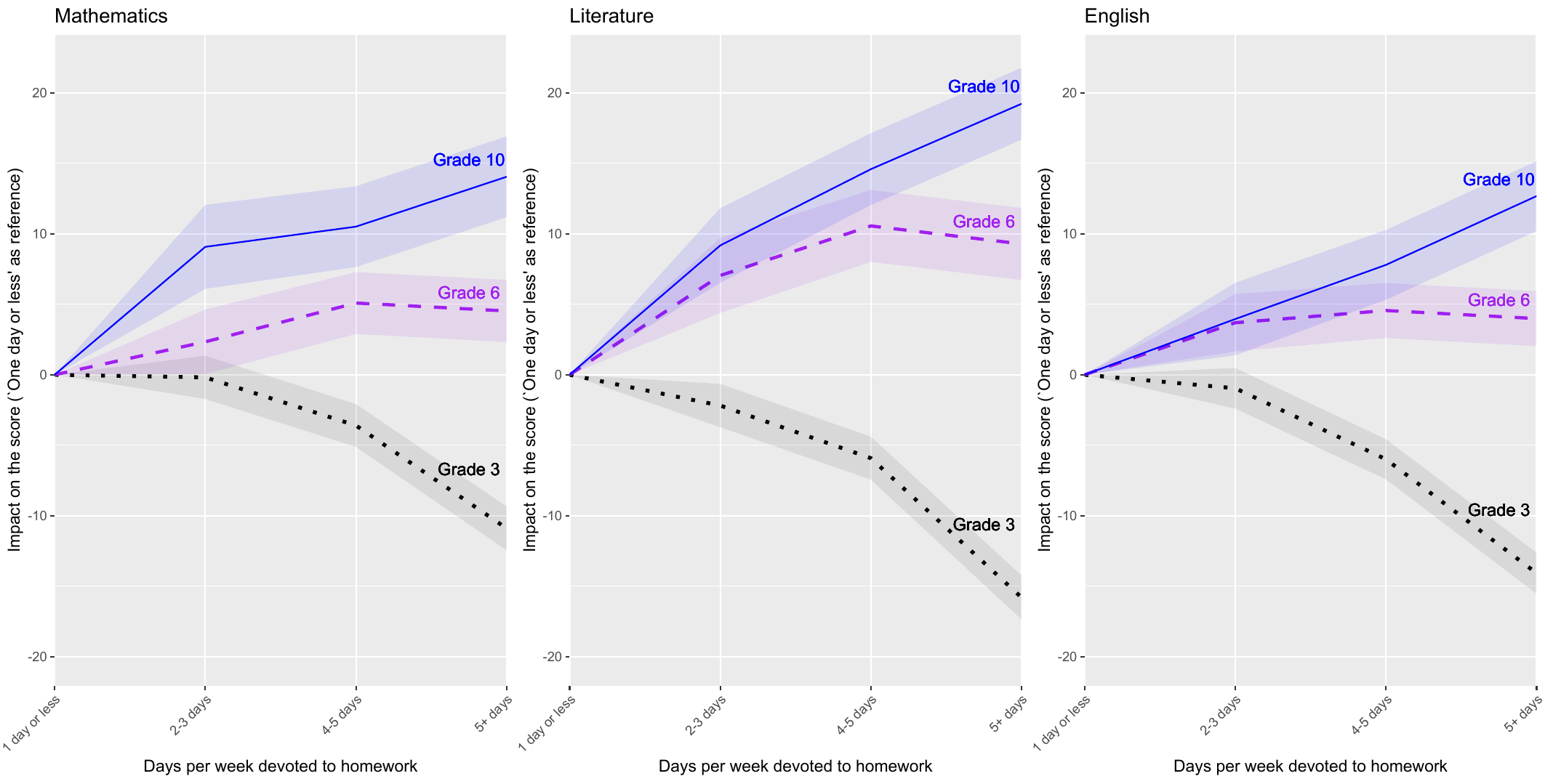}\\ 
\raggedright
\footnotesize{\textit{Reading}: In Grade 3 in mathematics, spending more that 5 days a week on homework, as opposed to one day or less, reduces a student's score by around 10 points (which represents 2\% of the normalised average score and 10\% of the normalised standard deviation, respectively equal to 500 and 100). $\pm$ standard deviation around the curve.}
\end{figure}

Estimates are summarised in Figure~\ref{stud.eff}. In terms of student's effort, a clear trend emerges, whatever the subject: in Grade 3, the more days per week devoted to homework, the lower the scores (compared with the lowest category, i.e. `one day or less'). In Grade 6, an increase in the number of days dedicated to homework goes with an increase in results up to 4-5 days per week, then decreases thereafter. In Grade 10, scores increase with the number of days dedicated to homework. 
The negative impact of the number of days dedicated to homework in the first grade seems to indicate a reverse causality (which could be confirmed by further investigations), i.e. it is poor school results that implies more time devoted to homework (the volume of homework being lower than in the higher grades). As students progress in their studies, effort seems to have an increasing impact, culminating in a clear positive impact in Grade 10. Finally, if we focus on the highest effort category (`five days or more') in Grade 10, we can see that the greatest positive impact is for literature (+19.23 points), then mathematics (+14.05 points), and finally English (+12.68 points).

The number of days per week dedicated to homework cannot be considered as a purely effort variable for the child, as homework is supervised by parents (particularly in early childhood). However, children gain independence as they progress through the grades and, as time goes on, the more work they do, the better their results. This form of emancipation seems to be weaker in English, compared with the other two subjects. 

\begin{figure}[!htp]
\centering
\caption{\label{stud.parent}Impact of parental investment on student’s score} \vspace{0.2cm}
\includegraphics[width=18cm]{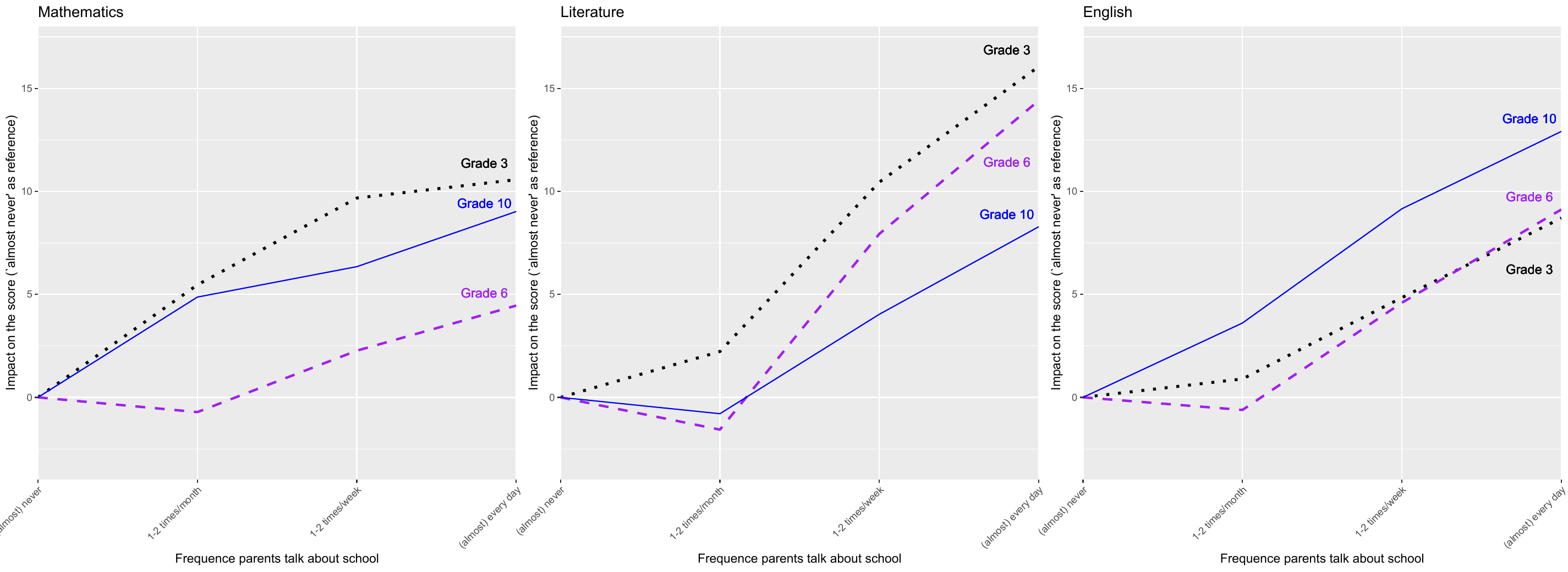}\\ 
\raggedright
\footnotesize{\textit{Reading}:  In Grade 10 in mathematics, if the parents talk to their child about school `once or twice a month' instead of `never', the child's score increases by around 5 points (which represents 1\% of the normalised average score and 5\% of the normalised standard deviation, respectively equal to 500 and 100).}
\end{figure}

The impact of parental investment on child's achievement also seems to go hand in hand with lower emancipation in English. First of all, we observe in Table~\ref{parent-invest} that, whatever the subject or grade (with some few exceptions), if the frequency with which parents talk to their child about school increases, then student achievement improves.  However, between Grades 3 and 10, if we focus on the highest category of investment (`every or almost every day'), the positive impact of parental investment decreases slightly in mathematics (from 10.6 to 9.0 points) and sharply in literature (from 16.1 to 8.3 points), while it rises sharply in English (from 8.7 to 12.9 points). Hence, between Grades 3 and 10, the impact of parental investment is clearly increasing in English, unlike in the the two other subjects. These results are shown in Figure~\ref{stud.parent} (where standard deviations have not been shown for the sake of clarity). 

Finally, if we look at all the explanatory variables in Grade 10 (Table~\ref{parent-invest}), we see that in mathematics and literature, the impact of child's effort exceeds the impact of parental investment and is comparable to the impact of parents' level of education. This is not the case in English: the impact of the child's effort and that of the parents' investment are roughly comparable, while the impact of parents' level of education is much greater.

% -------------------------------------------
% -------------------------------------------
\section{Adressing endogeneity: Instrumental variable (IV) analysis}
\label{sec-IV}
% -------------------------------------------
% -------------------------------------------

% -------------------------------------------
\subsection{Identification strategy}
\label{sec-identification}
% -------------------------------------------

Section~\ref{sec-OLS} relies on linear regressions to estimate the relationship between parental environment and student achievement. The advantage is that they allow for detailed analysis, as in Figure~\ref{Interactions} where, from Grade to Grade and by subject, we can see the differentiated impact of the three categories of parental education. However, these estimates may be biased due to the endogeneity of the dependent variables of interest. For instance, parents' level of education is likely correlated with unobserved family characteristics, such as motivation or inherited abilities, which also support a child's accumulation of human capital. This correlation can make it difficult to isolate the true causal effect of parental schooling. 

\begin{figure}[!htp]
\centering
\caption{\label{GG-all} Disparities in tertiary education between women and men in 1960} \vspace{0.3cm}
\small
\begin{tabular}{cc}
All the countries & Southwest zoom  \\ 
\includegraphics[width=8cm]{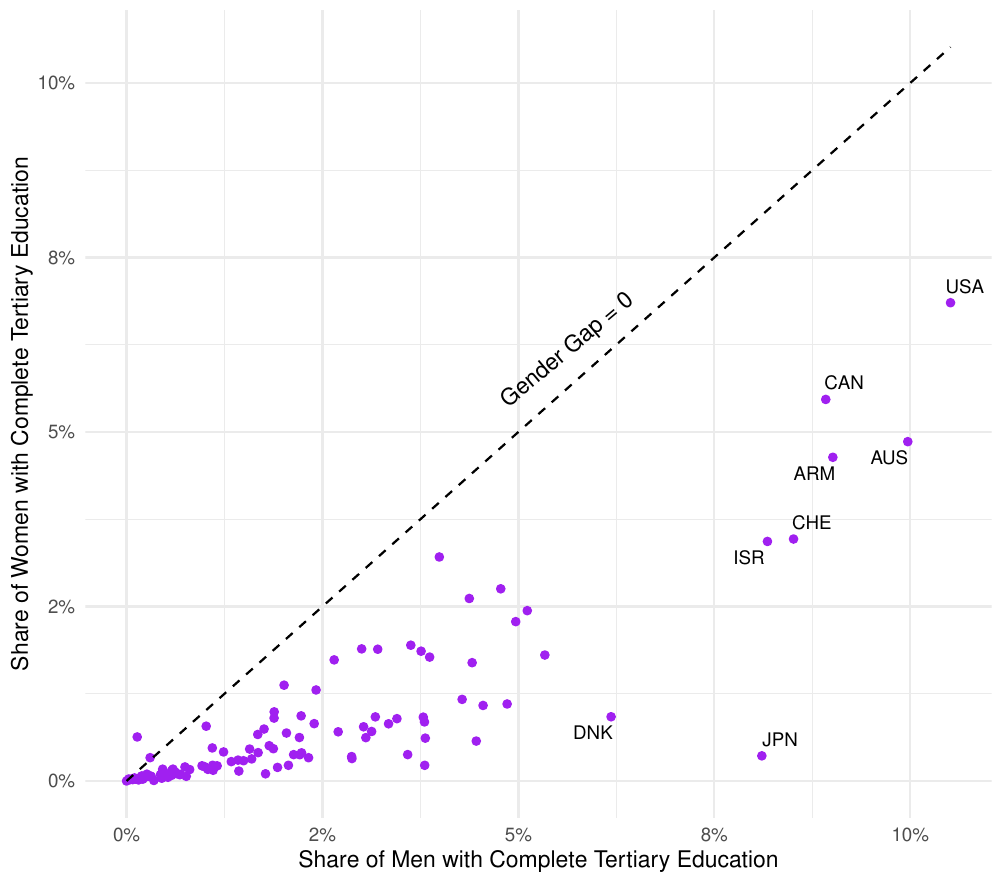} & %
\includegraphics[width=8cm]{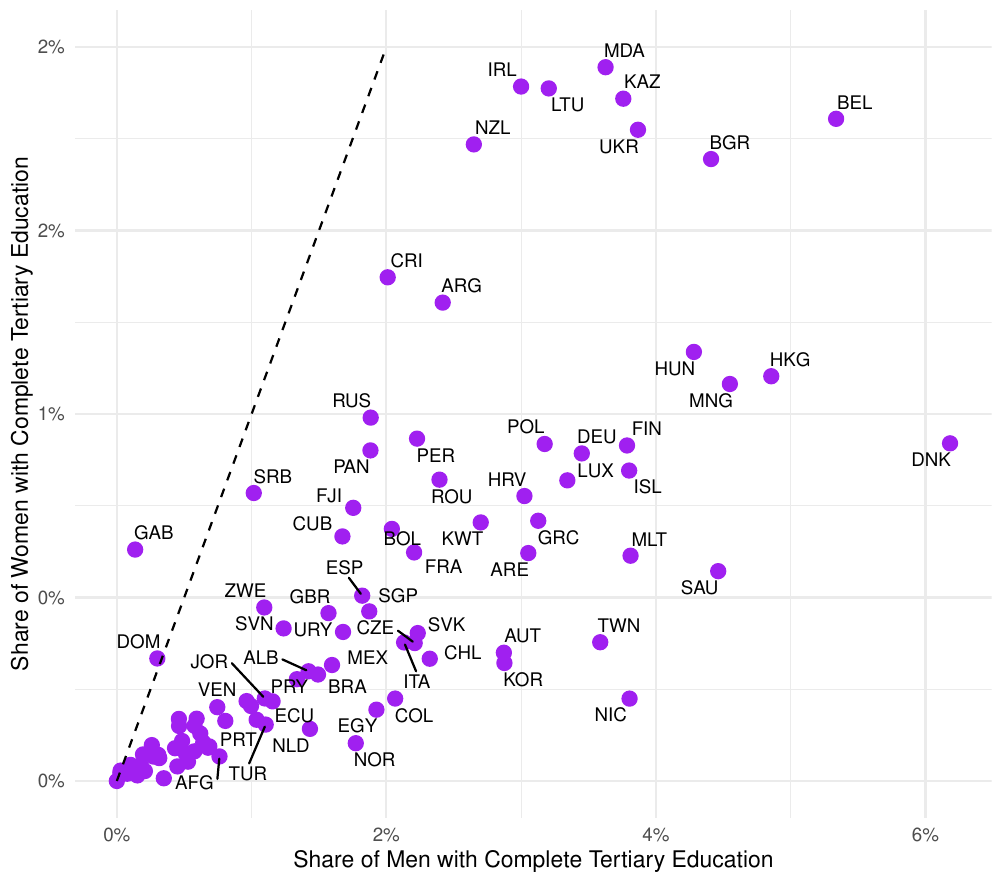} \\  
\end{tabular}
\raggedright
%\vspace{0.3cm}
\footnotesize{\textit{Reading}: In 1960, in Australia (AUS), approximately 5\% of adult women had completed higher education, compared to approximately 10\% of adult men.}\\ \vspace{0.2cm}
\footnotesize{\textit{Source}: Educational Attainment Data,  \cite{BL13}.}
\end{figure}

\begin{figure}[!htp]
\caption{\label{GG-curve} Gender gaps in tertiary education in 1960}
\vspace{0.2cm}
\centering
\includegraphics[width=10cm]{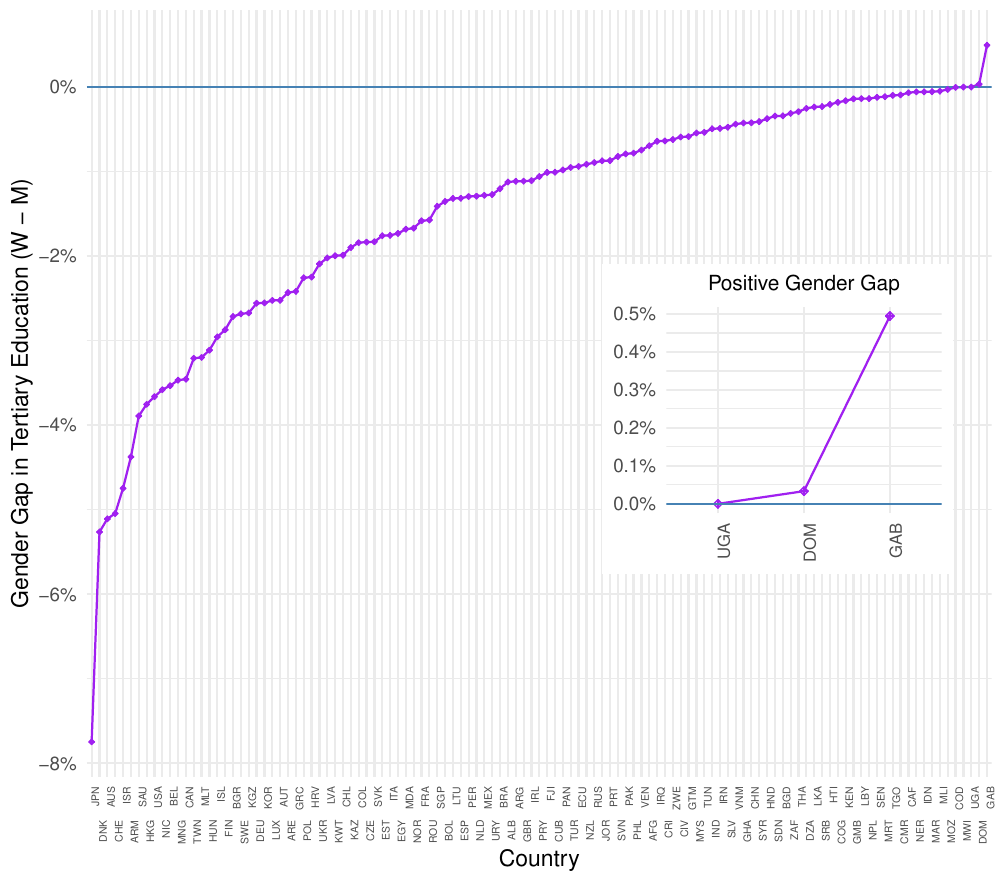}\\ 
\raggedright
\footnotesize{\textit{Reading}: In 1960, in Gabon (GAB), the share of women with complete tertiary education was 0.5\% higher than that of men.}\\ \vspace{0.2cm}
\footnotesize{\textit{Source}: Educational Attainment Data,  \cite{BL13}.}
\end{figure}

To address this endogeneity issue and identify the causal impact of parents' level of education on the academic performance of students, we employ an instrumental variable approach. Following \cite{CCS18}, who recommend the use of historical sources of exogenous variation, our chosen instrument is the gender gap in tertiary education in 1960 in the country of birth of the more educated parent. This variable is constructed using the Educational Attainment Data from \cite{BL13}. The gender gap is specifically calculated as the difference between the share of women and men who completed tertiary education. The viability of this instrument rests on its significant variation across the countries relevant to our analysis, a fact illustrated in Figures~\ref{GG-all} and~\ref{GG-curve}. The parents in our database come from 115 different countries, out of the 146 studied by \cite{BL13} in 1960. For these countries, Figure~\ref{GG-all} plots the share of women versus the share of men with complete tertiary education, demonstrating substantial cross-country differences. Figure~\ref{GG-curve} further details this variation by ranking countries according to the gender gap (`share of women' minus `share of men'), showing that the gap is negative for almost all the countries we are interested in, indicating lower rates of higher education for women.

Our identification strategy relies on the assumption that the gender gap in the parent's country of origin influences their child's academic performance exclusively through its effect on that parent's educational attainment. While this assumption cannot be tested directly, the richness of our data allows us to mitigate most concerns. For instance, one potential issue is that a country's wealth might be correlated with its gender gap, and that the wealth of the parents' country of origin could independently influence a child's academic success. The association between the gender gap in tertiary education and wealth is not confirmed in the literature. By analyzing global educational gender gaps (not only in tertiary education), \cite{MZKV19} show a relative weak correlation of 0.25 with economic growth, and that there is no significant difference when considering per capita income measures instead of growth. Focusing on the gender gap in tertiary education, \cite{HPP24} show that it can be negative or positive regardless of wealth, and establish that the difference between countries is partially explained by differences in social norms, relating to gender, within each country.\footnote{For example, the United States, which in 1960 was the richest country in the world, had the $9^{\textnormal{th}}$ most unequal gender gap in completed tertiary education in our database, among the 115 countries observed.}

The causal effect of the parent's highest level education on child's score is estimated using a two-stage least squares (2SLS) model. We first regress the endogenous variable, the parents' highest level of education (PHE, which can take three values, namely 0, 1 and 2 for, respectively, ISCED 0–2, 3–5, and 6–8), on the instrument, the gender gap in tertiary education (GG), including all the exogenous controls presented in regression~(\ref{first-reg}). As in Figure~\ref{GG-curve}, the gender gap is calculated as the difference between the share of women and the share of men who have completed higher education. The first-stage equation is:
\begin{equation} \label{IV-reg-first}
\textnormal{PHE}_{i} = \pi_0 + \pi_1 \textnormal{GG}_{i} + \textstyle\sum_{k} \sigma_k\,s_{ki} + \textstyle\sum_{k} \lambda_k\,h_{ki} + a_t + b_s + \nu_{its}\,.
\end{equation}

A negative value for $\pi_1$ can be expected, which would mean that the greater the gender inequality in education in the country of origin, the more educated parents tend to be. The literature explains this pattern through selective migration. For foreign parents, most of whom come from developing countries in our study, migration can be perceived as an opportunity.\footnote{By examining the Great Migration (1940–1970), during which millions of African Americans left the segregationist South for cities in the North, \cite{De22} challenges the notion that moving to areas offering better opportunities systematically guarantees greater intergenerational social mobility.} Because of mobility costs and spatial frictions, it is typically the most educated who are able to migrate \cite[][]{SSVN21, BCdLHKP24}. This pattern is, for instance, confirmed in \cite{BNS06} on the basis of three concrete case studies (Kerala in India, Bangladesh, and Albania).

We then use the predicted values of parents' highest level of education ($\widehat{\textnormal{PHE}}_{i}$) from the first stage to estimate the impact on student scores ($y_{its}$). The second-stage equation is:
\begin{equation} \label{IV-reg-second}
y_{its} = \alpha_0 + \alpha_1 \widehat{\textnormal{PHE}}_{i} + \textstyle\sum_{k} \beta_k\,s_{ki} + \textstyle\sum_{k} \gamma_k\,h_{ki} + a_t + b_s + \epsilon_{its}\,.
\end{equation}
In this framework, the coefficient $\alpha_1$ represents the causal effect of parents' highest level of education on student academic achievement. This regression is therefore equivalent to Equation~(\ref{first-reg}), but with an instrumented version of parents' highest level of education, and without it being put into dummy variables (which would have required a larger number of instruments, which could have posed potential problems in terms of validity).

As in Section~\ref{sec-OLS}, we complete the previous regression with interaction components between parental highest education and the student's grade level. Again, we compare the grades two by two (3 vs. 6, 6 vs. 10 and 3 vs. 10), running a regression for each possible grades pair $(u,v)$ where $u<v$, and excluding the non-concerned third grade. With $g$ indicating the grade, we introduce a dummy variable $I(u,v)$ which takes the value $0$ if $g=u$ (reference grade), and $1$ if $g=v$. When we compare the grades $u$ and $v$, we obtain the following expression, with $g \in \{u,v\}$:
\begin{equation} \label{IV-reg-second-interactions}
y_{igts} = \alpha_0 + \alpha_1 \widehat{\textnormal{PHE}}_{i} + \eta I(u,v) + \alpha_2 \widehat{\textnormal{PHE}}_i I(u,v) + \textstyle\sum_{k} \beta_k\,s_{ki} + \textstyle\sum_{k} \gamma_k\,h_{ki} + a_t + b_s + \epsilon_{igts}\,.
\end{equation}
This regression is the instrumented version of Equation~(\ref{second-reg}). Again the $\eta$ coefficient--which indicates the extra points obtained on average by the students in Grade $v$, compared with Grade $u$, whatever the parent group--is not informative due to the normalisation of the scores. The coefficient we are interest in is $\alpha_2$, which corresponds to the average impact, from Grade $u$ to Grade $v$, of a one-level increase in the highest level of parental education (among the three possible levels).

% -------------------------------------------
\subsection{Results}
\label{sec-IV_Results}
% -------------------------------------------

The results of the instrumental variable estimations of Equations~(\ref{IV-reg-first}) and~(\ref{IV-reg-second}) for the three subjects (mathematics, literature and English), aggregating all grade levels, are presented in Table~\ref{IV_all_subjects}.
\begin{table}[htbp]
\caption{Global and causal impact (IV) of parents' highest level of education on student's score}
\label{IV_all_subjects}
\vspace{0.2cm}
\centering
\footnotesize
\begin{tabular}{l ccc}
\toprule
 & {(1)} & {(2)} & {(3)} \\
 & {Mathematics} & {Literature} & {English} \\
\cmidrule(lr){1-1}\cmidrule(lr){2-2}\cmidrule(lr){3-3}\cmidrule(lr){4-4}
\multicolumn{4}{l}{\textbf{First stage: Impact of gender gap in tertiary education (GG)}} \\
\multicolumn{4}{l}{\textbf{on parents' highest level of education (PHE)}}\\ 
Gender gap in tertiary education & -0.051\sym{***} & -0.051\sym{***} & -0.052\sym{***} \\
 & (0.003) & (0.003) & (0.003) \\
F-test of excluded instrument & 248.06\sym{***} & 242.11\sym{***} & 250.91\sym{***} \\
 & & & \\
\multicolumn{4}{l}{\textbf{Reduced form: Impact of GG on student's score}} \\
Gender gap in tertiary education & -1.86\sym{***} & -2.634\sym{***} & -3.764\sym{***} \\
 & (0.533) & (0.568) & (0.487) \\
 & & & \\
\multicolumn{4}{l}{\textbf{Second stage: Impact of PHE, instrumented by GG, on  student's score}} \\
Parents highest education & 37.46\sym{***} & 51.30\sym{***} & 72.47\sym{***} \\
 & (10.46) & (11.36) & (9.89) \\
\cmidrule(lr){1-1}\cmidrule(lr){2-2}\cmidrule(lr){3-3}\cmidrule(lr){4-4}
Academic Year FE & Yes & Yes & Yes \\
School FE & Yes & Yes & Yes \\
Child's characteristics & Yes & Yes & Yes \\
Household's characteristics & Yes & Yes & Yes \\
\cmidrule(lr){1-1}\cmidrule(lr){2-2}\cmidrule(lr){3-3}\cmidrule(lr){4-4}
Observations & 297491 & 299350 & 297851 \\
\bottomrule
\multicolumn{4}{l}{\footnotesize Standard errors in parentheses}\\
\multicolumn{4}{l}{\footnotesize \sym{*} \(p<0.10\), \sym{**} \(p<0.05\), \sym{***} \(p<0.01\)}\\
\end{tabular}
\end{table}
The first-stage results show a strong and significant relationship between our instrument, the gender gap in tertiary education (GG), and the endogenous variable, parents' highest level of education (PHE). The coefficient on the gender gap is negative and statistically significant across all three subjects (approximately -0.05). This suggests that a larger gender gap in the parents' country of origin, which indicates lower educational attainment for women, is associated with a higher level of education for the parents in our sample, consistent with a selective migration pattern. The F-statistics for the excluded instrument are well above conventional thresholds for weak instruments, with values of 248.06 for mathematics, 242.11 for literature, and 250.91 for English, confirming the instrument's strength. The reduced-form estimates confirm a statistically significant negative relationship between the gender gap and children's scores across all subjects. 

The second-stage results in Table~\ref{IV_all_subjects} reveal a positive and significant causal effect of parents' highest education on their children's academic scores in all subjects. An increase in the parents' education level (recalling that we have three possible values, 0, 1 and 2) leads to a score increase of 37.46 points in mathematics, 51.30 points in literature, and 72.47 points in English. These results confirm that the parental environment has a substantial causal influence on student achievement after addressing potential endogeneity (persistent effect). These results also establish that the impact appears to be stronger in English, which confirms our observations in Section~\ref{sec-OLS}.

Table~\ref{IV_by_grade_reformatted} presents, still on the basis of Equations~(\ref{IV-reg-first}) and~(\ref{IV-reg-second}), the results disaggregated by grade level, allowing us to observe the evolution of this causal impact. In all grade-level regressions, the first-stage F-statistics are robust, ranging from 65.45 to 99.80, which supports the validity of the instrument across all subsamples. Now we focus on analyzing the results of the second-stage. One clear fact is the similar trend in mathematics and literature. The causal effect of parents' highest education is positive and significant in Grade 3 (at 1\%) and Grade 6 (at 5\%), but decreasing between the two grade levels. By Grade 10, the effect becomes statistically insignificant. In contrast to the other subjects, the causal effect of parents' education on English scores remains positive and statistically significant across all three grades (always at 1\%). While the magnitude of the coefficient decreases between Grade 6 and Grade 10, the influence of parental education persists strongly into adolescence for foreign language learning. As an illustration, in Grade 10, an increase in the parents' education level leads, on average, to an insignificant increase of about 10 points in mathematics and literature, and a significant increase of more than 50 points in English. This pattern seems to confirm that the causal influence of parental education on mathematics and literature performance diminishes as the child gets older, while it increases in English (Matthew effect), as established in Section~\ref{sec-OLS}.

%%%%%% TABLE %%%%%% 
\afterpage{
    \clearpage 
\begin{landscape}
\vspace*{1cm} 
\footnotesize 
\centering
\begin{longtable}{l*{9}{c}}
\caption{Marginal and causal impact (IV) of parents' highest level of education on student's score, by grade level}
\label{IV_by_grade_reformatted}\\
\toprule
 & \multicolumn{3}{c}{Mathematics} & \multicolumn{3}{c}{Literature} & \multicolumn{3}{c}{English} \\
\cmidrule(lr){2-4} \cmidrule(lr){5-7} \cmidrule(lr){8-10}
 & {(1)} & {(2)} & {(3)} & {(4)} & {(5)} & {(6)} & {(7)} & {(8)} & {(9)} \\
 & {Gr.3} & {Gr.6} & {Gr.10} & {Gr.3} & {Gr.6} & {Gr.10} & {Gr.3} & {Gr.6} & {Gr.10} \\
\cmidrule(lr){1-1}\cmidrule(lr){2-4}\cmidrule(lr){5-7}\cmidrule(lr){8-10}
\endfirsthead
\caption{Impact of Parental Education on student's core (continued)}\\
\toprule
 & \multicolumn{3}{c}{Mathematics} & \multicolumn{3}{c}{Literature} & \multicolumn{3}{c}{English} \\
\cmidrule(lr){2-4} \cmidrule(lr){5-7} \cmidrule(lr){8-10}
 & {(1)} & {(2)} & {(3)} & {(4)} & {(5)} & {(6)} & {(7)} & {(8)} & {(9)} \\
 & {Gr.3} & {Gr.6} & {Gr.10} & {Gr.3} & {Gr.6} & {Gr.10} & {Gr.3} & {Gr.6} & {Gr.10} \\
\cmidrule(lr){1-1}\cmidrule(lr){2-4}\cmidrule(lr){5-7}\cmidrule(lr){8-10}
\endhead
\bottomrule
\multicolumn{10}{r}{\textit{Continued on next page}} \\
\endfoot
\bottomrule
\multicolumn{10}{l}{\footnotesize Standard errors in parentheses}\\
\multicolumn{10}{l}{\footnotesize \sym{*} \(p<0.10\), \sym{**} \(p<0.05\), \sym{***} \(p<0.01\)}\\
\endlastfoot
% --- TABLE BODY ---
\multicolumn{10}{l}{\textbf{First stage: Impact of gender gap in tertiary education (GG) on parents' highest level of education (PHE)}} \\
Gender gap in tertiary education & -0.040\sym{***} & -0.057\sym{***} & -0.065\sym{***} & -0.040\sym{***} & -0.056\sym{***} & -0.063\sym{***} & -0.041\sym{***} & -0.057\sym{***} & -0.066\sym{***} \\
 & (0.005) & (0.006) & (0.007) & (0.005) & (0.006) & (0.007) & (0.005) & (0.006) & (0.007) \\
F-test of excluded instrument & 65.45\sym{***} & 99.60\sym{***} & 90.25\sym{***} & 66.91\sym{***} & 97.22\sym{***} & 84.09\sym{***} & 67.24\sym{***} & 99.80\sym{***} & 91.58\sym{***} \\
 & & & & & & & & & \\
\multicolumn{10}{l}{\textbf{Reduced form: Impact of GG in Education on student's score}} \\
Gender gap in tertiary education & -2.763\sym{***} & -1.924\sym{**} & -0.859 & -3.706\sym{***} & -2.492\sym{**} & -0.601 & -3.229\sym{***} & -4.619\sym{***} & -3.609\sym{***} \\
 & (0.790) & (0.896) & (1.109) & (0.795) & (1.029) & (0.981) & (0.743) & (0.802) & (0.968) \\
 & & & & & & & & & \\
\multicolumn{10}{l}{\textbf{Second stage: Impact of PHE, instrumented by GG in Ed., on student's score}} \\
Parents highest education & 72.65\sym{***} & 33.65\sym{**} & 13.40 & 93.22\sym{***} & 41.84\sym{**} & 11.03 & 82.58\sym{***} & 80.15\sym{***} & 54.05\sym{***} \\
 & (21.04) & (15.94) & (17.13) & (21.83) & (18.57) & (15.65) & (19.87) & (15.27) & (15.00) \\
\cmidrule(lr){1-1}\cmidrule(lr){2-4}\cmidrule(lr){5-7}\cmidrule(lr){8-10}
Academic Year FE & Yes & Yes & Yes & Yes & Yes & Yes & Yes & Yes & Yes \\
School FE & Yes & Yes & Yes & Yes & Yes & Yes & Yes & Yes & Yes \\
Child's characteristics & Yes & Yes & Yes & Yes & Yes & Yes & Yes & Yes & Yes \\
Household's characteristics & Yes & Yes & Yes & Yes & Yes & Yes & Yes & Yes & Yes \\
\cmidrule(lr){1-1}\cmidrule(lr){2-4}\cmidrule(lr){5-7}\cmidrule(lr){8-10}
Observations & 133592 & 113824 & 51761 & 133899 & 113876 & 51575 & 132695 & 113472 & 51684 \\
\end{longtable}
\end{landscape}
    \clearpage 
}

\begin{figure}[!htp]
\centering
\caption{\label{Interactions-IV}Marginal and causal impact (IV) of parents’ highest level of education on student's score} \vspace{0.2cm}
\includegraphics[width=18cm]{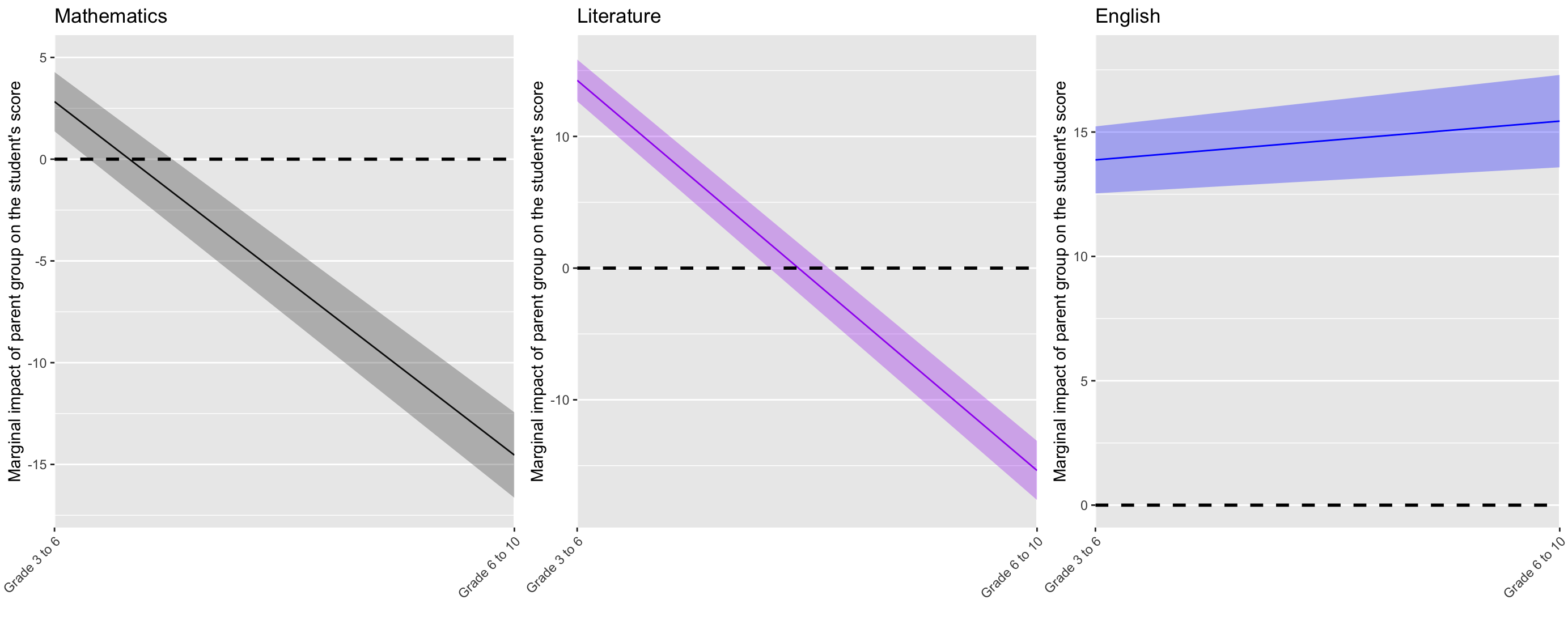}\\ 
\raggedright
\footnotesize{\textit{Reading}: Between Grades 3 and 6 in literature, increasing the parents' highest level of education (among the three possible categories: ISCED 0–2, ISCED 3–5 and ISCED 6–8) by one category increases the score by an average of 14.26 points. Thereafter, between Grade 6 and Grade 10 in literature, the marginal impact of a higher category for the parents' highest level of education, on the student's score, is reduced by an average of 15.37 points. $\pm$ standard deviation around the line.}
\end{figure}

We now discuss the estimation results of Equation~(\ref{IV-reg-second-interactions}), which includes an interaction term between parental highest education and the student's grade level. These results are summarised in Tables~\ref{G3-G6-IV}, \ref{G6-G10-IV} and~\ref{G3-G10-IV} for, respectively, the transition from Grade 3 to Grade 6, Grade 6 to Grade 10, and Grade 3 to Grade 10. Once again, the causal effect of the variable `parents' highest level of education' remains positive and significant, regardless of the subject and regardless of the econometric specification. We are now interested in the marginal impact from one grade to another, considering only estimates that include all control variables. These results can be viewed in Figure~\ref{Interactions-IV}.

In mathematics, between Grades 3 and 6, a one-category increase in the parent's level of education leads, on average, to a slight increase (significant at only 10\%) of 2.8 points of the student's score. Then, between Grades 6 and 10, this impact is significantly negative (-14.5 points). The trend is similar in literature, except that the positive impact is significant between Grades 3 and 6 (+14.3 points). As in mathematics, it is then negative and significant between Grades 6 and 10 (-15.4 points). The results for these two subjects confirm a general downward trend in the impact of parents' level of education on student's score in mathematics between the ages of 8 and 15, and a bell-shaped trend in literature. The pattern in English is completely different. The marginal impact remains strongly positive (and significant) between Grades 3 and 6 (+13.9 points), and Grades 6 and 10 (+15.5 points). Overall, between Grades 3 and 10, the marginal impact decreases in mathematics and literature (by -4.7 and -2.6 points, respectively) while it increases by 32.5 points in English.

To sum up, these instrumental variable estimates reinforce the findings from our initial OLS models, providing stronger causal evidence for a mixed Matthew effect: the influence of the parental environment diminishes for mathematics and literature as students mature (with a bell-shaped curve in literature), but it is unambiguously amplified in foreign language acquisition.

% -------------------------------------------
% -------------------------------------------
\section{Discussion, related literature, and policy implications}
\label{sec-discuss}
% -------------------------------------------
% -------------------------------------------

\noindent \textit{\textbf{Matthew effect in education}}. Few empirical studies have attempted to test the Matthew effect hypothesis in education, most of them applied to reading abilities and its impact on acquisition of literacy (and other related skills). They all agree that the differences in reading abilities in the early education stages continue on until adulthood \cite[]{CS97, Ri10}, but the results are mixed on the existence of a Matthew effect: There is not a strong support for a pattern of widening or decreasing achievement differences \cite[]{Pfetal14}. Some papers find that the effect is strongly increasing \cite[][]{AB95, Ho01} , others that it is intermediate \cite[][]{BR98} and some that it is not even significant or related to social background \cite[][]{Shetal95, Pretal11}. But no study has tested, on a unified database, the existence of a possible Matthew effect on the core subjects of mathematics, literature and the main foreign language (in our case English).

While our study confirms a persistent impact of parental environment on child's academic performance (a fact widely accepted since the Coleman report, 1966), the results differ between subjects as regards the Matthew effect. From age 11 to 15, the effect of parent's level of education decreases in mathematics and literature, while it increases in English. At age of 15, spending `5 days or more' doing homework (compared with `one day or less') increases child achievement, but more strongly in literature than in mathematics, and more strongly in mathematics than in English. Similarly, at the same age, the impact of `the frequency with which parents talk to their child about school' is stronger in English, followed by mathematics and then by literature (although the order is reversed at age 8).

To sum up, these results therefore reflect a partial emancipation (from the influence of parental environment) in mathematics and literature, while social determinism increases in English. In the first two subjects, the results echo the work of cognitive psychology initiated by Jean Piaget, according to which the child is partly master of his or her own development: They have an intrinsic ability to learn, without this necessarily being transmitted by others, and their strategies and involvement play a role in their academic performance \cite[][]{OAN00}. The notion of an age of consent (for responsible choice) can therefore make sense, including in education, and this age can be set between 12 and 16, as proposed by \cite{HPRU17}.

Conversely, our results confirm the social dimension of learning a foreign language, compared with other academic subjects. In that case, external factors appear to play a decisive role in the learning process \cite[][]{Vy78}. In addition, the theory of \cite{CH07} which describes a path-dependency in the formation of cognitive skills, seems to be confirmed. As \cite{Ga68} and \cite{Do98} point out, the acquisition of a new language involves a great deal of integrative motivation (in the sense that people are interested in learning a language because they want to communicate with the other language community), and parents play a crucial role in encouraging this integrative motivation (as opposed to instrumental motivation). According to \cite{Ga68}, parents play two roles in their child's success in learning a second language: an active role which consists of actively and consciously encouraging their child to learn the language, and a (more important) passive role, which consists of the attitudes that parents have towards the community whose language their child is learning. 

\noindent \textit{\textbf{Policy implications}}. Our results, which need to be confirmed using other databases and complementary methodologies, have several implications for educational policies. The first concerns remediation programmes, aimed at improving the skills of children experiencing difficulties. The second concerns national selection processes in higher education (`Grandes Ecoles' in France, for instance), which include foreign language skills as a criterion for admission.

With regard to remedial programmes, the main recommendation resulting from the empirical estimates of \cite{CH07} and \cite{CHS10} is to focus (adolescent) remediation strategies for disadvantaged children on the development of non-cognitive skills. Our results indicate that programmes focusing on cognitive skills can also be effective, particularly in mathematics. These results are in line with those of \cite{Ba07,Ba08}, who assesses US postsecondary remediation programmes. He first observes that the degree of deficiency (depth) and the number of deficient basic skill areas (breadth) are good predictors of successful math remediation: Those who require the least remediation are the most likely to remediate successfully. But he also observes that when remediation works (at a low rate, unfortunately) it works extremely well: `students who remediate successfully in mathematics exhibit attainment that is comparable to that of students who achieve college mathematics skill without the need for remediation' \cite[][Page 442]{Ba08}.

The second implication in terms of public education policy concerns the weight of foreign languages in the selection process to access higher education, at different levels. We have found that parents' education and involvement are essential factors in children's success in learning a foreign language. What's more, the child's effort has (slightly) less impact on results than in subjects such as mathematics or literature. This is a strong sign of inequality of opportunity, and including foreign languages as an admission criterion reinforces this inequality. It therefore seems essential that educational systems give a high priority to foreign language teaching, particularly in non-English-speaking countries, from an early age, so as not to further penalise children from socially disadvantaged backgrounds.

% -------------------------------------------
\bibliographystyle{ecca}
\bibliography{202307_aymeric-lavaine-magdalou-V13}

\balance 

% -------------------------------------------
\newpage
\singlespacing
\onecolumn
\appendix
% -------------------------------------------
% -------------------------------------------
\section{Appendix}
% -------------------------------------------
% -------------------------------------------

%
\begin{figure}[!htp]
\caption{\label{Lit-Scores} CDFs of literature scores according to parents' highest level of education} \vspace{0.2cm}
\small
\begin{tabular}{cc}
\includegraphics[width=8cm]{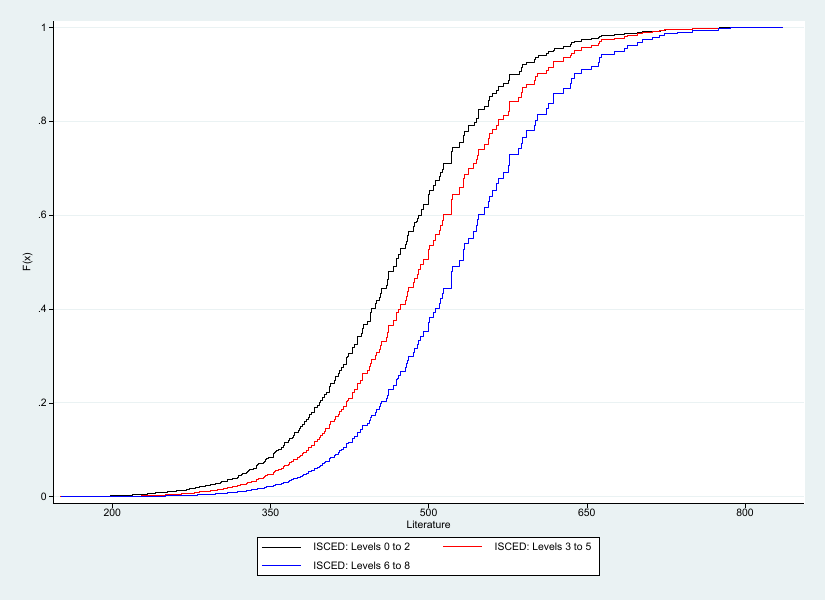} & %
\includegraphics[width=8cm]{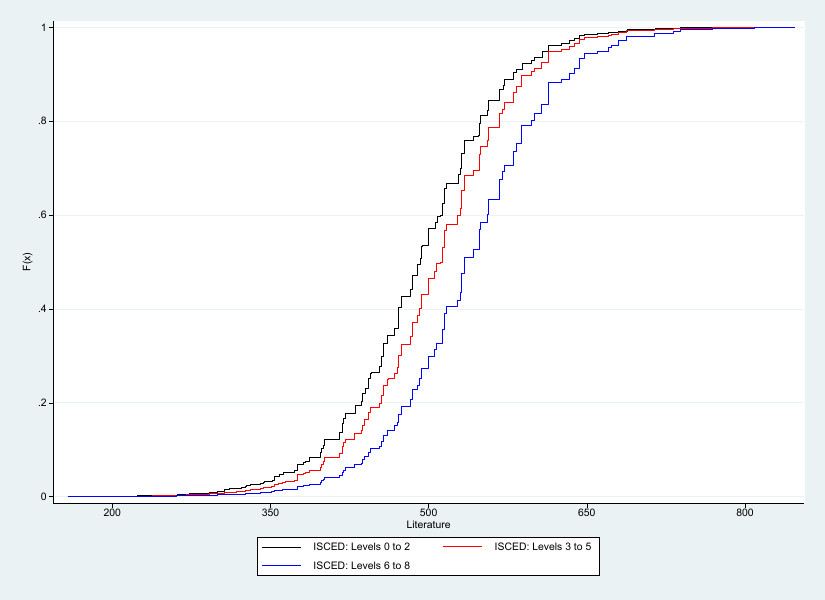} \\ 
Grade 3 & Grade 10 \\  
\end{tabular}

\end{figure} 

\vspace{1cm}

\begin{figure}[!htp] \label{}
\caption{\label{Eng-Scores} CDFs of English scores according to parents' highest level of education} \vspace{0.2cm}
\centering
\small
\begin{tabular}{cc}
\includegraphics[width=8cm]{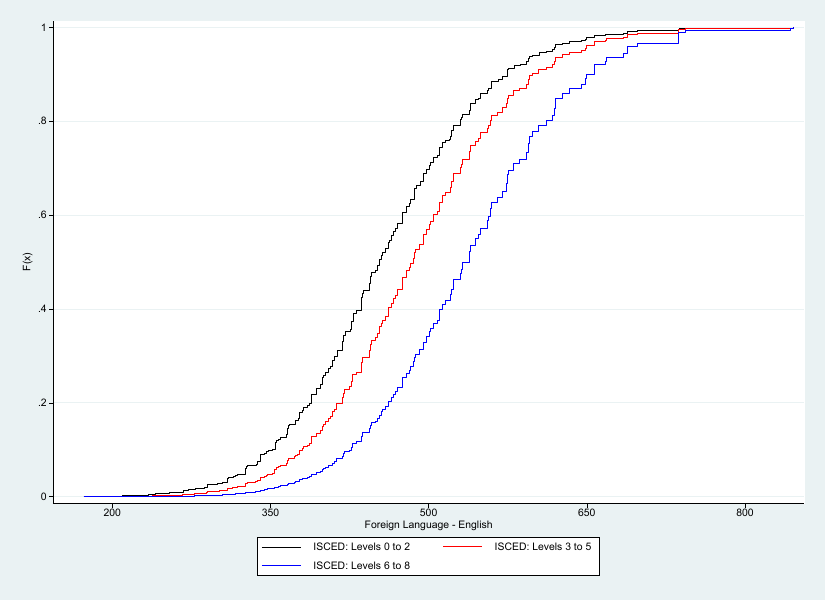} & %
\includegraphics[width=8cm]{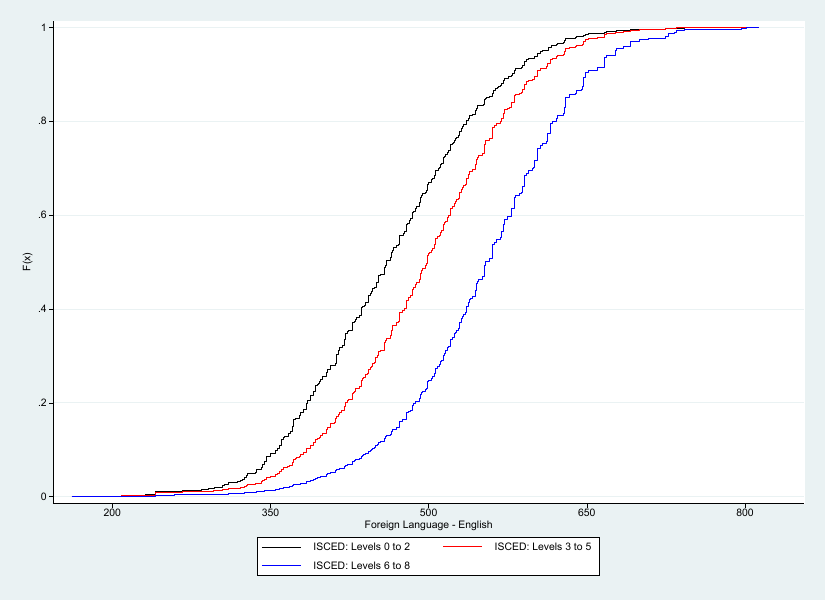} \\ 
Grade 3 & Grade 10 \\  
\end{tabular}
\end{figure} 
\newpage

\footnotesize

%%%%%% TABLE %%%%%% 
\def\sym#1{\ifmmode^{#1}\else\(^{#1}\)\fi}
\renewcommand{\arraystretch}{0.65}

 \vspace{0.2cm}
  \centering
  \footnotesize
\begin{longtable}{llll}
  \caption[]{Variables Description\label{var-desc}}\\
\toprule
      	&\multicolumn{1}{l}{Type}&\multicolumn{1}{l}{Values}&\multicolumn{1}{l}{Description}\\      
\cmidrule(lr){1-1} \cmidrule(lr){2-2} \cmidrule(lr){3-3} \cmidrule(lr){4-4}
\endfirsthead
\caption[]{Variables Description (continued)}\\
\toprule
			&\multicolumn{1}{l}{Type}&\multicolumn{1}{l}{Values}&\multicolumn{1}{l}{Description}\\ 
\cmidrule(lr){1-1} \cmidrule(lr){2-2} \cmidrule(lr){3-3} \cmidrule(lr){4-4}
\endhead
\hline
\endfoot
\hline
\endlastfoot
\multicolumn{4}{l}{\textbf{Students marks}}\\
\cmidrule(lr){1-1} \cmidrule(lr){2-2} \cmidrule(lr){3-3} \cmidrule(lr){4-4}
Mathematics &  Num.&	0 - 1034&  Transformed Mathematics mark\\
Literature  &   Num.&	0 - 1001&  Transformed Literature mark\\
English (Foreign Language)  &  Num.&	0 - 845&  Transformed English mark\\ \\
\multicolumn{4}{l}{\textbf{Parents' highest education}}\\
\cmidrule(lr){1-1} \cmidrule(lr){2-2} \cmidrule(lr){3-3} \cmidrule(lr){4-4}
Parents' highest education&  Qual.&	0 - 2&  0 ISCED 0 to 2 \\
 &  &	&  1 ISCED 3 to 5\\
\bottomrule
\end{longtable}

%%%%%% TABLE %%%%%% 
\begin{table}[htbp]\centering
\caption{Summary Statistics\label{tab2}}
 \vspace{0.2cm}
  \centering
  \footnotesize
\begin{tabular}{l*{1}{llll}}
\toprule
        &    Mean&     Stand. Dev.\\
\cmidrule(lr){1-1} \cmidrule(lr){2-2} \cmidrule(lr){3-3}
\multicolumn{2}{l}{\textbf{Students marks}}\\
\cmidrule(lr){1-1} \cmidrule(lr){2-2} \cmidrule(lr){3-3}
Mathematics &  517.4546&  95.00513\\
Literature  &   505.935&  101.8927\\
English Foreign Language  &  517.1402&   93.7307\\
\multicolumn{2}{l}{\textbf{Parents' highest education}}\\
\cmidrule(lr){1-1} \cmidrule(lr){2-2} \cmidrule(lr){3-3}
Parents highest education&  1.494245&  .6642053\\
\multicolumn{2}{l}{\textbf{Child's characteristics}}\\
\cmidrule(lr){1-1} \cmidrule(lr){2-2} \cmidrule(lr){3-3}
Child's country of birth   &  1.044064&  .2052369\\
Days/week dedicated to homework    &  3.310194&  .7581023\\
Child's gender &  1.502942&  .4999921\\
\multicolumn{2}{l}{\textbf{Household's characteristics}}\\
\cmidrule(lr){1-1} \cmidrule(lr){2-2} \cmidrule(lr){3-3}
Mother's country of birth   &  1.165642&  .3717597\\
Father's country of birth   &  1.159382&  .3660328\\
Freq. books used at home   &  3.375672&   .837836\\
Freq. computer used at home   &  3.794766&  .5249957\\
Freq. internet used at home   &  3.865501&  .4511002\\
Nb of books at home    &  3.515174&  1.203279\\
Mother's labour situation   &  1.651739&  1.015354\\
Father's labour situation   &  1.303231&  .7056803\\
Freq. parents talk about school with child   &   3.82281&  .5012083\\
Freq. parents schedule homework with child   &  3.476625&  .9059994\\
Freq. parents help for homework   &   2.99286&  1.098742\\
Freq. parents check homework with child   &  3.440297&  .9775864\\
\hline
\(N\)    &   327163&      &      &      \\
\bottomrule
\end{tabular}
\end{table}

%%%%%% TABLE %%%%%% 
\begin{table}[htbp]\centering
\caption[]{Average scores according to grade level\label{tab3}}
\vspace{0.2cm}
  \centering
  \footnotesize
\begin{tabular}{L{3cm}C{2cm}C{2cm}C{2cm}}
\toprule
      &{(1)}&{(2)}&{(3)}\\
      &{Gr.3}&{Gr.6}&{Gr.10}\\
\cmidrule(lr){1-1}\cmidrule(lr){2-2}\cmidrule(lr){3-3}\cmidrule(lr){4-4}
Mathematics		&   513.29&   521.52&   519.33\\

Literature &   514.21&   489.35&   522.26\\

English	  &   515.02&   515.77&   526.17\\

\cmidrule(lr){1-1}\cmidrule(lr){2-2}\cmidrule(lr){3-3}\cmidrule(lr){4-4}
\(N\)    &   145096&   123811&    52637\\
\bottomrule
\end{tabular}
\end{table}

%%%%%% TABLE %%%%%% 
%\begin{landscape}
\begin{table}[htbp]\centering
\caption[]{Average scores according to grade level, by academic year\label{tab4}}
\vspace{0.2cm}
  \centering
  \footnotesize
\begin{tabular}{l*{11}{c}}
\toprule
     &{(1)}&{(2)}&{(3)}&{(4)}&{(5)}&{(6)}&{(7)}&{(8)}&{(9)}&{(10)}&{(11)}\\
     &{2016}&{2016}&{2017}&{2017}&{2017}&{2018}&{2018}&{2018}&{2019}&{2019}&{2019}\\
     &{Gr.3}&{Gr.6}&{Gr.3}&{Gr.6}&{Gr.10}&{Gr.3}&{Gr.6}&{Gr.10}&{Gr.3}&{Gr.6}&{Gr.10}\\
\cmidrule(lr){1-1}\cmidrule(lr){2-2}\cmidrule(lr){3-3}\cmidrule(lr){4-4}\cmidrule(lr){5-5}\cmidrule(lr){6-6}\cmidrule(lr){7-7}\cmidrule(lr){8-8}\cmidrule(lr){9-9}\cmidrule(lr){10-10}\cmidrule(lr){11-11}\cmidrule(lr){12-12}
Mathematics&  507.1&  509.8&  514.5&  515.3&  514.1&  518.1&  545.5&  522.9&  515.9&  518.4&  520.9\\
     
Literature &  510.5&  436.3&  513.3&  514.8&  518.7&  517.6&  524.1&  526.5&  517.0&  516.6&  521.1\\
     
English &  509.2&  510.0&  515.3&  516.1&  522.2&  520.1&  520.1&  530.7&  517.7&  520.5&  525.1\\
    
\cmidrule(lr){1-1}\cmidrule(lr){2-2}\cmidrule(lr){3-3}\cmidrule(lr){4-4}\cmidrule(lr){5-5}\cmidrule(lr){6-6}\cmidrule(lr){7-7}\cmidrule(lr){8-8}\cmidrule(lr){9-9}\cmidrule(lr){10-10}\cmidrule(lr){11-11}\cmidrule(lr){12-12}
\(N\)   &  47061&  44249&  32390&  24578&  18029&  36535&  31060&  19594&  29110&  23924&  15014\\
\bottomrule
\end{tabular}
\end{table}
%\end{landscape}

%%%%%% TABLE %%%%%% 
%\begin{landscape}
\begin{table}[htbp]\centering
\caption{Students distribution according to parent highest education, grade level and academic year\label{stud-dist}}
\vspace{0.2cm}
  \centering
  \footnotesize
\begin{tabular}{l*{11}{c}}
\toprule
     &{(1)}&{(2)}&{(3)}&{(4)}&{(5)}&{(6)}&{(7)}&{(8)}&{(9)}&{(10)}&{(11)}\\
     &{2016}&{2016}&{2017}&{2017}&{2017}&{2018}&{2018}&{2018}&{2019}&{2019}&{2019}\\
     &{Gr.3}&{Gr.6}&{Gr.3}&{Gr.6}&{Gr.10}&{Gr.3}&{Gr.6}&{Gr.10}&{Gr.3}&{Gr.6}&{Gr.10}\\
\cmidrule(lr){1-1}\cmidrule(lr){2-2}\cmidrule(lr){3-3}\cmidrule(lr){4-4}\cmidrule(lr){5-5}\cmidrule(lr){6-6}\cmidrule(lr){7-7}\cmidrule(lr){8-8}\cmidrule(lr){9-9}\cmidrule(lr){10-10}\cmidrule(lr){11-11}\cmidrule(lr){12-12}
ISCED 0--2     &  11.15$\%$&  10.90$\%$&  7.945$\%$&  9.520$\%$&  9.335$\%$&  8.845$\%$&  10.15$\%$&  8.574$\%$&  8.737$\%$&  9.296$\%$&  8.468$\%$\\
     
ISCED 3--5     &  33.40$\%$&  35.43$\%$&  29.87$\%$&  32.93$\%$&  35.53$\%$&  28.87$\%$&  31.07$\%$&  32.08$\%$&  26.72$\%$&  28.57$\%$&  30.14$\%$\\
     
ISCED 6--8     &  55.45$\%$&  53.67$\%$&  62.19$\%$&  57.55$\%$&  55.14$\%$&  62.28$\%$&  58.79$\%$&  59.35$\%$&  64.54$\%$&  62.13$\%$&  61.40$\%$\\
    
\cmidrule(lr){1-1}\cmidrule(lr){2-2}\cmidrule(lr){3-3}\cmidrule(lr){4-4}\cmidrule(lr){5-5}\cmidrule(lr){6-6}\cmidrule(lr){7-7}\cmidrule(lr){8-8}\cmidrule(lr){9-9}\cmidrule(lr){10-10}\cmidrule(lr){11-11}\cmidrule(lr){12-12}
Total   &   100$\%$&   100$\%$&   100$\%$&   100$\%$&   100$\%$&   100$\%$&   100$\%$&   100$\%$&   100$\%$&   100$\%$&   100$\%$\\
\cmidrule(lr){1-1}\cmidrule(lr){2-2}\cmidrule(lr){3-3}\cmidrule(lr){4-4}\cmidrule(lr){5-5}\cmidrule(lr){6-6}\cmidrule(lr){7-7}\cmidrule(lr){8-8}\cmidrule(lr){9-9}\cmidrule(lr){10-10}\cmidrule(lr){11-11}\cmidrule(lr){12-12}
\(N\)   &  47081&  44305&  33077&  24706&  18425&  36619&  31036&  19909&  29769&  24644&  15470\\
\bottomrule
\end{tabular}
\end{table}
%\end{landscape}

%%%%%% TABLE %%%%%% 
\begin{table}[htbp]
\caption{Impact of parents' highest level of education -- Mathematics\label{maths-first}}
\vspace{0.2cm}
  \centering
    \footnotesize
\begin{tabular}{L{6.5cm}C{2.5cm}C{2.5cm}C{2.5cm}}
\toprule
          &{(1)}&{(2)}&{(3)}\\
          &{Mathematics}&{Mathematics}&{Mathematics}\\
\cmidrule(lr){1-1}\cmidrule(lr){2-2}\cmidrule(lr){3-3}\cmidrule(lr){4-4}
\multicolumn{4}{l}{\textbf{Parents' highest education}}\\
ISCED 0--2   &      0     &      0     &      0     \\
          &     (.)     &     (.)     &     (.)     \\
ISCED 3--5 &    15.81\sym{***}&    16.72\sym{***}&    10.27\sym{***}\\
          &   (0.479)     &   (0.517)     &   (0.589)     \\
ISCED 6--8  &    41.40\sym{***}&    42.92\sym{***}&    29.70\sym{***}\\
          &   (0.491)     &   (0.530)     &   (0.625)     \\
\multicolumn{4}{l}{\textbf{Constant}}\\
          &    482.3\sym{***}&    486.6\sym{***}&    476.5\sym{***}\\
          &   (0.499)     &   (1.150)     &   (2.661)     \\
\cmidrule(lr){1-1}\cmidrule(lr){2-2}\cmidrule(lr){3-3}\cmidrule(lr){4-4}
Academic year fixed-effect	&    Yes&    Yes&    Yes\\
School fixed-effect		&    Yes&    Yes&    Yes\\
Child's characteristics	&    No&    Yes&    Yes\\
Household's characteristics	&    No&    No&    Yes\\
\cmidrule(lr){1-1}\cmidrule(lr){2-2}\cmidrule(lr){3-3}\cmidrule(lr){4-4}
Observations    &   415226     &   367492     &   316294     \\
Adjusted \(R^{2}\) &    0.160     &    0.172     &    0.204     \\
\bottomrule
\multicolumn{4}{l}{\footnotesize Standard errors in parentheses}\\
\multicolumn{4}{l}{\footnotesize \sym{*} \(p<0.10\), \sym{**} \(p<0.05\), \sym{***} \(p<0.01\)}\\
\end{tabular}
\end{table}

%%%%%% TABLE %%%%%% 

\begin{table}[htbp]
\caption{Impact of parents' highest level of education -- Literature\label{lit-first}}
\vspace{0.2cm}
  \centering
    \footnotesize
\begin{tabular}{L{6.5cm}C{2.5cm}C{2.5cm}C{2.5cm}}
\toprule
           &{(1)}&{(2)}&{(3)}\\
          &{Literature}&{Literature}&{Literature}\\
\cmidrule(lr){1-1}\cmidrule(lr){2-2}\cmidrule(lr){3-3}\cmidrule(lr){4-4}
\multicolumn{4}{l}{\textbf{Parents' highest education}}\\
ISCED 0--2     &      0     &      0     &      0     \\
          &     (.)     &     (.)     &     (.)     \\
ISCED 3--5 &    19.25\sym{***}&    20.27\sym{***}&    14.28\sym{***}\\
          &   (0.515)     &   (0.556)     &   (0.634)     \\
ISCED 6--8   &    46.42\sym{***}&    48.65\sym{***}&    35.89\sym{***}\\
          &   (0.527)     &   (0.569)     &   (0.673)     \\
\multicolumn{4}{l}{\textbf{Constant}}\\
          &    478.3\sym{***}&    504.4\sym{***}&    488.8\sym{***}\\
          &   (0.536)     &   (1.237)     &   (2.865)     \\
\cmidrule(lr){1-1}\cmidrule(lr){2-2}\cmidrule(lr){3-3}\cmidrule(lr){4-4} 
Academic year fixed-effect	&    Yes&    Yes&    Yes\\
School fixed-effect		&    Yes&    Yes&    Yes\\
Child's characteristics	&    No&    Yes&    Yes\\
Household's characteristics	&    No&    No&    Yes\\          
\cmidrule(lr){1-1}\cmidrule(lr){2-2}\cmidrule(lr){3-3}\cmidrule(lr){4-4}
Observations    &   414805     &   367757     &   316529     \\
Adjusted \(R^{2}\) &    0.146     &    0.175     &    0.197     \\
\bottomrule
\multicolumn{4}{l}{\footnotesize Standard errors in parentheses}\\
\multicolumn{4}{l}{\footnotesize \sym{*} \(p<0.10\), \sym{**} \(p<0.05\), \sym{***} \(p<0.01\)}\\
\end{tabular}
\end{table}

%%%%%% TABLE %%%%%% 
\begin{table}[htbp]
\caption{Impact of parents' highest level of education -- English\label{eng-first}}
\vspace{0.2cm}
  \centering
    \footnotesize
\begin{tabular}{L{6.5cm}C{2.5cm}C{2.5cm}C{2.5cm}}
\toprule
           &{(1)}&{(2)}&{(3)}\\
          &{English}&{English}&{English}\\
\cmidrule(lr){1-1}\cmidrule(lr){2-2}\cmidrule(lr){3-3}\cmidrule(lr){4-4}        
\multicolumn{4}{l}{\textbf{Parents' highest education}}\\
ISCED 0--2     &      0     &      0     &      0     \\
          &     (.)     &     (.)     &     (.)     \\
ISCED 3--5 &    21.81\sym{***}&    21.75\sym{***}&    16.06\sym{***}\\
          &   (0.437)     &   (0.471)     &   (0.535)     \\
ISCED 6--8   &    52.55\sym{***}&    52.34\sym{***}&    39.68\sym{***}\\
          &   (0.448)     &   (0.483)     &   (0.568)     \\
 \multicolumn{4}{l}{\textbf{Constant}}\\
          &    473.7\sym{***}&    492.3\sym{***}&    468.3\sym{***}\\
          &   (0.455)     &   (1.048)     &   (2.418)     \\
\cmidrule(lr){1-1}\cmidrule(lr){2-2}\cmidrule(lr){3-3}\cmidrule(lr){4-4} 
Academic year fixed-effect	&    Yes&    Yes&    Yes\\
School fixed-effect		&    Yes&    Yes&    Yes\\
Child's characteristics	&    No&    Yes&    Yes\\
Household's characteristics	&    No&    No&    Yes\\
\cmidrule(lr){1-1}\cmidrule(lr){2-2}\cmidrule(lr){3-3}\cmidrule(lr){4-4}
Observations    &   413466     &   365774     &   314863     \\
Adjusted \(R^{2}\) &    0.293     &    0.303     &    0.329     \\
\bottomrule
\multicolumn{4}{l}{\footnotesize Standard errors in parentheses}\\
\multicolumn{4}{l}{\footnotesize \sym{*} \(p<0.10\), \sym{**} \(p<0.05\), \sym{***} \(p<0.01\)}\\
\end{tabular}
\end{table}

%%%%%% TABLE : Grade 3 vs Grade 6 %%%%%%
\begin{landscape}
  \vspace{0.2cm}
  \centering
  \footnotesize
  \begin{longtable}{l*{12}{c}} % Nombre de colonnes ajusté à 12
  \caption[]{Impact of parents' highest level of education according to grade level -- \textbf{Grade 3 vs Grade 6} \label{G3-G6}}\\
  \toprule
  & \multicolumn{3}{c}{Mathematics} & \multicolumn{3}{c}{Literature} & \multicolumn{3}{c}{English} \\
  \cmidrule(lr){2-4} \cmidrule(lr){5-7} \cmidrule(lr){8-10}
  &{(1)}&{(2)}&{(3)}&{(1)}&{(2)}&{(3)}&{(1)}&{(2)}&{(3)}\\
  &{Math.}&{Math.}&{Math.}&{Lit.}&{Lit.}&{Lit.}&{Engl.}&{Engl.}&{Engl.}\\
  \cmidrule(lr){1-1}\cmidrule(lr){2-4}\cmidrule(lr){5-7}\cmidrule(lr){8-10}
  \endfirsthead
  \caption[]{Parents highest education's impact according to grade level (compared two-by-two) -- Mathematics, Literature, and English (continued)}\\
  \toprule
  & \multicolumn{3}{c}{Mathematics} & \multicolumn{3}{c}{Literature} & \multicolumn{3}{c}{English} \\
  \cmidrule(lr){2-4} \cmidrule(lr){5-7} \cmidrule(lr){8-10}
  & \multicolumn{3}{c}{Grade 3 vs Grade 6} & \multicolumn{3}{c}{Grade 3 vs Grade 10} & \multicolumn{3}{c}{Grade 6 vs Grade 10} \\
  \cmidrule(lr){2-4} \cmidrule(lr){5-7} \cmidrule(lr){8-10}
  &{(1)}&{(2)}&{(3)}&{(4)}&{(5)}&{(6)}&{(7)}&{(8)}&{(9)}\\
  &{Math}&{Math}&{Math}&{Lit.}&{Lit.}&{Lit.}&{Engl}&{Engl}&{Engl}\\
  \cmidrule(lr){1-1}\cmidrule(lr){2-4}\cmidrule(lr){5-7}\cmidrule(lr){8-10}
  \endhead
  \hline
  \endfoot
  \endlastfoot
  \multicolumn{10}{l}{\textbf{Parents highest education}}\\
  ISCED~0--2   & 0 & 0 & 0 & 0 & 0 & 0 & 0 & 0 & 0 \\
          & (.) & (.) & (.) & (.) & (.) & (.) & (.) & (.) & (.) \\
  ISCED~3--5   & 18.84\sym{***} & 19.05\sym{***} & 12.05\sym{***} & 19.91\sym{***} & 20.39\sym{***} & 13.30\sym{***} & 18.59\sym{***} & 19.22\sym{***} & 13.35\sym{***} \\
          & (0.757) & (0.762) & (0.864) & (0.845) & (0.840) & (0.950) & (0.698) & (0.700) & (0.794) \\
  ISCED~6--8   & 46.16\sym{***} & 46.16\sym{***} & 30.22\sym{***} & 44.50\sym{***} & 45.18\sym{***} & 28.44\sym{***} & 46.57\sym{***} & 47.07\sym{***} & 33.11\sym{***} \\
          & (0.745) & (0.750) & (0.874) & (0.832) & (0.826) & (0.960) & (0.687) & (0.689) & (0.803) \\
  \multicolumn{10}{l}{\textbf{Grade 3 vs Grade 6}}\\
\end{longtable}
\end{landscape}

%%%%%% TABLE : Grade 6 vs Grade 10 %%%%%%
\begin{landscape}
  \vspace{0.2cm}
  \centering
  \footnotesize
  \begin{longtable}{l*{12}{c}}
  \caption[]{Impact of parents' highest level of education according to grade level -- \textbf{Grade 6 vs Grade 10}}\label{G6-G10}\\
  \toprule
  & \multicolumn{3}{c}{Mathematics} & \multicolumn{3}{c}{Literature} & \multicolumn{3}{c}{English} \\
  \cmidrule(lr){2-4} \cmidrule(lr){5-7} \cmidrule(lr){8-10}
  &{(4)}&{(5)}&{(6)}&{(4)}&{(5)}&{(6)}&{(4)}&{(5)}&{(6)}\\
  &{Math.}&{Math.}&{Math.}&{Lit.}&{Lit.}&{Lit.}&{Engl.}&{Engl.}&{Engl.}\\
  \cmidrule(lr){1-1}\cmidrule(lr){2-4}\cmidrule(lr){5-7}\cmidrule(lr){8-10}
  \endfirsthead
  \caption[]{Parents highest education's impact according to grade level (compared two-by-two) -- \textbf{Grade 6 vs Grade 10} (continued)}\\
  \toprule
  & \multicolumn{3}{c}{Grade 6 vs Grade 10 - Math} & \multicolumn{3}{c}{Grade 3 vs Grade 6 - Lit.} & \multicolumn{3}{c}{Grade 3 vs Grade 6 - Eng.} \\
  \cmidrule(lr){2-4} \cmidrule(lr){5-7} \cmidrule(lr){8-10}
  &{(5)}&{(6)}&{(7)}&{(5)}&{(6)}&{(7)}&{(4)}&{(5)}&{(6)}\\
  &{Math}&{Math}&{Math}&{Lit.}&{Lit.}&{Lit.}&{Eng.}&{Eng.}&{Eng.}\\
  \cmidrule(lr){1-1}\cmidrule(lr){2-4}\cmidrule(lr){5-7}\cmidrule(lr){8-10}
  \endhead
  \hline
  \endfoot
  \endlastfoot
  \multicolumn{12}{l}{\textbf{Parents highest education}}\\
  ISCED~0--2   & 0 & 0 & 0 & 0 & 0 & 0 & 0 & 0 & 0 \\
          & (.) & (.) & (.) & (.) & (.) & (.) & (.) & (.) & (.) \\
  ISCED~3--5   & 18.40\sym{***} & 17.86\sym{***} & 11.03\sym{***} & 21.82\sym{***} & 22.05\sym{***} & 14.96\sym{***} & 22.10\sym{***} & 22.29\sym{***} & 15.87\sym{***} \\
          & (0.819) & (0.816) & (0.921) & (0.877) & (0.884) & (0.993) & (0.727) & (0.719) & (0.806) \\
  ISCED~6--8   & 47.02\sym{***} & 46.62\sym{***} & 31.28\sym{***} & 52.57\sym{***} & 53.50\sym{***} & 36.83\sym{***} & 51.79\sym{***} & 52.93\sym{***} & 37.69\sym{***} \\
          & (0.833) & (0.833) & (0.966) & (0.892) & (0.902) & (1.042) & (0.739) & (0.734) & (0.846) \\
  \multicolumn{12}{l}{\textbf{Grade 6 vs Grade 10}}\\
\end{longtable}
\end{landscape}

%%%%%% TABLE : Grade 3 vs Grade 10 %%%%%%
\begin{landscape}
  \vspace{0.2cm}
  \centering
  \footnotesize
  \begin{longtable}{l*{9}{c}}
  \caption[]{Impact of parents' highest level of education according to grade level -- \textbf{Grade 3 vs Grade 10}}\label{G3-G10}\\
  \toprule
  & \multicolumn{3}{c}{Mathematics} & \multicolumn{3}{c}{Literature} & \multicolumn{3}{c}{English} \\
  \cmidrule(lr){2-4} \cmidrule(lr){5-7} \cmidrule(lr){8-10}
  &{(7)}&{(8)}&{(9)}&{(7)}&{(8)}&{(9)}&{(7)}&{(8)}&{(9)}\\
  &{Math.}&{Math.}&{Math.}&{Lit.}&{Lit.}&{Lit.}&{Engl.}&{Engl.}&{Engl.}\\
  \cmidrule(lr){1-1}\cmidrule(lr){2-4}\cmidrule(lr){5-7}\cmidrule(lr){8-10}
  \endfirsthead
  \caption[]{Parents highest education's impact according to grade level (compared two-by-two) -- Mathematics, Literature, and English (continued)}\\
  \toprule
  & \multicolumn{3}{c}{Mathematics} & \multicolumn{3}{c}{Literature} & \multicolumn{3}{c}{English} \\
  \cmidrule(lr){2-4} \cmidrule(lr){5-7} \cmidrule(lr){8-10}
  &{(7)}&{(8)}&{(9)}&{(7)}&{(8)}&{(9)}&{(7)}&{(8)}&{(9)}\\
  &{Math}&{Math}&{Math}&{Lit.}&{Lit.}&{Lit.}&{Engl}&{Engl}&{Engl}\\
  \cmidrule(lr){1-1}\cmidrule(lr){2-4}\cmidrule(lr){5-7}\cmidrule(lr){8-10}
  \endhead
  \hline
  \endfoot
  \endlastfoot
  \multicolumn{10}{l}{\textbf{Parents' highest education}}\\
  ISCED~0--2   & 0 & 0 & 0 & 0 & 0 & 0 & 0 & 0 & 0 \\
          & (.) & (.) & (.) & (.) & (.) & (.) & (.) & (.) & (.) \\
  ISCED~3--5   & 18.97\sym{***} & 19.00\sym{***} & 12.33\sym{***} & 21.19\sym{***} & 21.90\sym{***} & 15.32\sym{***} & 19.19\sym{***} & 19.93\sym{***} & 14.05\sym{***} \\
          & (0.769) & (0.765) & (0.877) & (0.751) & (0.745) & (0.850) & (0.711) & (0.708) & (0.806) \\
  ISCED~6--8   & 45.24\sym{***} & 45.39\sym{***} & 30.08\sym{***} & 46.60\sym{***} & 48.01\sym{***} & 32.75\sym{***} & 45.51\sym{***} & 47.29\sym{***} & 33.20\sym{***} \\
          & (0.777) & (0.776) & (0.917) & (0.758) & (0.754) & (0.888) & (0.718) & (0.717) & (0.843) \\
  \multicolumn{10}{l}{\textbf{Grade 3 vs Grade 10}}\\
\end{longtable}
\end{landscape}

%%%%%% TABLE %%%%%% 
\begin{landscape}
\vspace{0.2cm}
  \centering
    \footnotesize
\begin{longtable}{l*{9}{c}}
\caption{Impact of parental investment and student's effort \label{parent-invest}}\\
\toprule
   & \multicolumn{3}{c}{Mathematics} & \multicolumn{3}{c}{Literature} & \multicolumn{3}{c}{English} \\
  \cmidrule(lr){2-4} \cmidrule(lr){5-7} \cmidrule(lr){8-10}
   &{(1)}&{(2)}&{(3)}&{(4)}&{(5)}&{(6)}&{(7)}&{(8)}&{(9)}\\
%          &{Math}&{Math}&{Math}&{Lit.}&{Lit.}&{Lit.}&{Engl}&{Engl}&{Engl}\\
          &{Gr.3}&{Gr.6}&{Gr.10}&{Gr.3}&{Gr.6}&{Gr.10}&{Gr.3}&{Gr.6}&{Gr.10}\\
\cmidrule(lr){1-1}\cmidrule(lr){2-4}\cmidrule(lr){5-7}\cmidrule(lr){8-10}

\endfirsthead
\caption{Parental investment and child effort (continued) }\\
\toprule
 					&{(1)}&{(2)}&{(3)}&{(4)}&{(5)}&{(6)}&{(7)}&{(8)}&{(9)}\\
          &{Math}&{Math}&{Math}&{Lit.}&{Lit.}&{Lit.}&{Engl}&{Engl}&{Engl}\\
          &{Gr.3}&{Gr.6}&{Gr.10}&{Gr.3}&{Gr.6}&{Gr.10}&{Gr.3}&{Gr.6}&{Gr.10}\\
\cmidrule(lr){1-1}\cmidrule(lr){2-4}\cmidrule(lr){5-7}\cmidrule(lr){8-10}
\endhead
\hline
\endfoot
\endlastfoot
\multicolumn{10}{l}{\textbf{Parents' highest education}}\\
ISCED 0--2         &      0     &      0     &      0     &      0     &      0     &      0     &      0     &      0     &      0     \\
          &     (.)     &     (.)     &     (.)     &     (.)     &     (.)     &     (.)     &     (.)     &     (.)     &     (.)     \\
ISCED 3--5 &    12.04\sym{***}&    10.99\sym{***}&    3.304\sym{**} &    15.27\sym{***}&    14.68\sym{***}&    7.688\sym{***}&    14.71\sym{***}&    16.40\sym{***}&    16.46\sym{***}\\
          &   (0.861)     &   (0.909)     &   (1.496)     &   (0.864)     &   (1.054)     &   (1.325)     &   (0.806)     &   (0.806)     &   (1.290)     \\
ISCED 6--8   &    30.01\sym{***}&    30.84\sym{***}&    17.90\sym{***}&    32.99\sym{***}&    35.96\sym{***}&    24.54\sym{***}&    35.62\sym{***}&    39.08\sym{***}&    42.99\sym{***}\\
          &   (0.914)     &   (0.972)     &   (1.584)     &   (0.917)     &   (1.128)     &   (1.403)     &   (0.855)     &   (0.862)     &   (1.366)     \\
\multicolumn{10}{l}{\textbf{Days/week dedicated to homework}}\\
\end{longtable}
\end{landscape}

%%%%%% TABLE %%%%%% 
{\setlength{\tabcolsep}{1.5pt}
\begin{landscape}
\def\sym#1{\ifmmode^{#1}\else\(^{#1}\)\fi}
\begin{longtable}{l*{9}{D{.}{.}{-1}}}
\caption{Impact of parent's highest level of education (predicted values by first stage IV regression) according to grade level - \textbf{Grade 3 vs Grade 6}\label{G3-G6-IV}}\\
\toprule\endfirsthead\hline\endhead\hline\endfoot\endlastfoot
          &\multicolumn{3}{c}{Mathematics}&\multicolumn{3}{c}{Literature}&\multicolumn{3}{c}{English}\\
          \cmidrule(lr){2-4}\cmidrule(lr){5-7} \cmidrule(lr){8-10}
           &\multicolumn{1}{c}{(1)}&\multicolumn{1}{c}{(2)}&\multicolumn{1}{c}{(3)}&\multicolumn{1}{c}{(1)}&\multicolumn{1}{c}{(2)}&\multicolumn{1}{c}{(3)}&\multicolumn{1}{c}{(1)}&\multicolumn{1}{c}{(2)}&\multicolumn{1}{c}{(3)}\\
          &\multicolumn{1}{c}{Math.}&\multicolumn{1}{c}{Math.}&\multicolumn{1}{c}{Math.}&\multicolumn{1}{c}{Lit.}&\multicolumn{1}{c}{Lit.}&\multicolumn{1}{c}{Lit.}&\multicolumn{1}{c}{Engl.}&\multicolumn{1}{c}{Engl.}&\multicolumn{1}{c}{Engl.}\\
          \cmidrule(lr){1-1}\cmidrule(lr){2-4}\cmidrule(lr){5-7} \cmidrule(lr){8-10}
\multicolumn{10}{l}{\textbf{Parent's highest education}} \\ 
  &    67.52\sym{***}&    65.63\sym{***}&    46.97\sym{***}&    63.02\sym{***}&    65.58\sym{***}&    59.24\sym{***}&    57.75\sym{***}&    58.80\sym{***}&    69.05\sym{***}\\
          &   (1.062)     &   (1.058)     &   (11.74)     &   (1.171)     &   (1.151)     &   (12.82)     &   (0.985)     &   (0.976)     &   (10.85)     \\
\multicolumn{10}{l}{\textbf{Grade 3 vs Grade 6}} \\          
I(3,6)   &    9.682\sym{***}&    8.517\sym{***}&   -8.106\sym{***}&   -36.42\sym{***}&   -36.62\sym{***}&   -53.68\sym{***}&   -15.35\sym{***}&   -16.14\sym{***}&   -28.49\sym{***}\\
\end{longtable}
\end{landscape}
}

%%%%%% TABLE %%%%%% 
{\setlength{\tabcolsep}{1.5pt}
\begin{landscape}
\def\sym#1{\ifmmode^{#1}\else\(^{#1}\)\fi}
\begin{longtable}{l*{9}{D{.}{.}{-1}}}
\caption{Impact of parent's highest level of education (predicted values by first stage IV regression) according to grade level - \textbf{Grade 6 vs Grade 10}\label{G6-G10-IV}}\\
\toprule\endfirsthead\hline\endhead\hline\endfoot\endlastfoot
                   &\multicolumn{3}{c}{Mathematics}&\multicolumn{3}{c}{Literature}&\multicolumn{3}{c}{English}\\
                    \cmidrule(lr){2-4}\cmidrule(lr){5-7} \cmidrule(lr){8-10}
                     &\multicolumn{1}{c}{(4)}&\multicolumn{1}{c}{(5)}&\multicolumn{1}{c}{(6)}&\multicolumn{1}{c}{(4)}&\multicolumn{1}{c}{(5)}&\multicolumn{1}{c}{(6)}&\multicolumn{1}{c}{(4)}&\multicolumn{1}{c}{(5)}&\multicolumn{1}{c}{(6)}\\
                    &\multicolumn{1}{c}{Math.}&\multicolumn{1}{c}{Math.}&\multicolumn{1}{c}{Math.}&\multicolumn{1}{c}{Lit.}&\multicolumn{1}{c}{Lit.}&\multicolumn{1}{c}{Lit.}&\multicolumn{1}{c}{Engl.}&\multicolumn{1}{c}{Engl.}&\multicolumn{1}{c}{Engl.}\\
                    \cmidrule(lr){1-1}\cmidrule(lr){2-4}\cmidrule(lr){5-7} \cmidrule(lr){8-10}
\multicolumn{10}{l}{\textbf{Parent's highest education}} \\                     
   &       66.80\sym{***}&       66.13\sym{***}&       34.14\sym{**} &       70.74\sym{***}&       73.56\sym{***}&       40.30\sym{***}&       65.08\sym{***}&       66.82\sym{***}&       76.60\sym{***}\\
                    &     (1.212)         &     (1.206)         &     (13.74)         &     (1.305)         &     (1.284)         &     (14.66)         &     (1.078)         &     (1.069)         &     (12.14)         \\
\multicolumn{10}{l}{\textbf{Grade 6 vs Grade 10}} \\                    
I(6,10)      &       22.26\sym{***}&       25.73\sym{***}&       1.693         &       35.62\sym{***}&       37.89\sym{***}&       16.24\sym{***}&      -10.08\sym{***}&      -8.706\sym{***}&      -29.77\sym{***}\\
\end{longtable}
\end{landscape}
}

%%%%%% TABLE %%%%%% 
{\setlength{\tabcolsep}{1.5pt}
\begin{landscape}
\def\sym#1{\ifmmode^{#1}\else\(^{#1}\)\fi}
\begin{longtable}{l*{9}{D{.}{.}{-1}}}
\caption{Impact of parent's highest level of education (predicted values by first stage IV regression) according to grade level - \textbf{Grade 3 vs Grade 10}\label{G3-G10-IV}}\\
\toprule\endfirsthead\hline\endhead\hline\endfoot\endlastfoot
                   &\multicolumn{3}{c}{Mathematics}&\multicolumn{3}{c}{Literature}&\multicolumn{3}{c}{English}\\
                    \cmidrule(lr){2-4}\cmidrule(lr){5-7} \cmidrule(lr){8-10}
                     &\multicolumn{1}{c}{(7)}&\multicolumn{1}{c}{(8)}&\multicolumn{1}{c}{(9)}&\multicolumn{1}{c}{(7)}&\multicolumn{1}{c}{(8)}&\multicolumn{1}{c}{(9)}&\multicolumn{1}{c}{(7)}&\multicolumn{1}{c}{(8)}&\multicolumn{1}{c}{(9)}\\
                    &\multicolumn{1}{c}{Math.}&\multicolumn{1}{c}{Math.}&\multicolumn{1}{c}{Math.}&\multicolumn{1}{c}{Lit.}&\multicolumn{1}{c}{Lit.}&\multicolumn{1}{c}{Lit.}&\multicolumn{1}{c}{Engl.}&\multicolumn{1}{c}{Engl.}&\multicolumn{1}{c}{Engl.}\\
                    \cmidrule(lr){1-1}\cmidrule(lr){2-4}\cmidrule(lr){5-7} \cmidrule(lr){8-10}
\multicolumn{10}{l}{\textbf{Parent's highest education}} \\
   &       65.68\sym{***}&       63.94\sym{***}&       36.96\sym{***}&       63.71\sym{***}&       65.60\sym{***}&       48.73\sym{***}&       57.45\sym{***}&       58.61\sym{***}&       50.93\sym{***}\\
                    &     (1.101)         &     (1.097)         &     (12.70)         &     (1.076)         &     (1.064)         &     (12.33)         &     (1.022)         &     (1.016)         &     (11.72)         \\
\multicolumn{10}{l}{\textbf{Grade 3 vs Grade 10}} \\                    
I(3,10)     &       19.91\sym{***}&       20.19\sym{***}&      -16.85\sym{***}&       18.21\sym{***}&       17.95\sym{***}&      -11.83\sym{***}&      -34.30\sym{***}&      -35.37\sym{***}&      -67.77\sym{***}\\
\end{longtable}
\end{landscape}
}

\newpage

\end{document}